\documentclass[aps,prb,twocolumn,superscriptaddress,floatfix]{revtex4-1}

\usepackage{graphicx}
\usepackage{dcolumn}
\usepackage{bm}
\usepackage{times}
\usepackage{color}

\begin{document}
\bibliographystyle{apsrev4-1}
\title{Magneto-transport and Electronic Structures of BaZnBi$_2$}
\author{Yi-Yan Wang}\thanks{These authors contributed equally to this paper}
\author{Peng-Jie Guo}\thanks{These authors contributed equally to this paper}
\author{Qiao-He Yu}
\author{Sheng Xu}
\author{Kai Liu}
\author{Tian-Long Xia}\email{tlxia@ruc.edu.cn}
\affiliation{Department of Physics, Renmin University of China, Beijing 100872, P. R. China}
\affiliation{Beijing Key Laboratory of Opto-electronic Functional Materials $\&$ Micro-nano Devices, Renmin University of China, Beijing 100872, P. R. China}
\date{\today}
\begin{abstract}
We report the magneto-transport properties and electronic structures of BaZnBi$_2$. BaZnBi$_2$ is a quasi-two-dimensional (2D) material with metallic behavior. The transverse magnetoresistance (MR) depends on magnetic field linearly and exhibits Shubnikov-de Haas (SdH) oscillation at low temperature and high field. The observed linear MR may originate from the disorder in samples or the edge conductivity in compensated two-component systems. The first-principles calculations reveal the absence of stable gapless Dirac fermion. Combining with the trivial Berry phase extracted from SdH oscillation, BaZnBi$_2$ is suggested to be a topologically trivial semimetal. Nearly compensated electron-like Fermi surfaces (FSs) and hole-like FSs coexist in BaZnBi$_2$.
\end{abstract}
\maketitle

\setlength{\parindent}{1em}
\section{Introduction}

Topological semimetals, including Dirac/Weyl/nodal line semimetals, have recently attracted much attention for their extraordinary physical properties\cite{PhysRevB.83.205101,wehling2014dirac,PhysRevB.85.195320,weng2016topological,liu2014discovery,neupane2014observationCd3As2,YLChen2014stableCd3As2,PhysRevLett.113.027603,liang2015ultrahigh,PhysRevX.5.011029,xu2015discoveryTaAs,huang2015weyl,PhysRevX.5.031013DingHongTaAs,NatPhysDingHongTaAs,yang2015weyl,xu2015discoveryNbAs,xu2015experimental,xu2016observation,liu2016evolution,arnold2016negative,zhang2016signatures,PhysRevX.5.031023,hu2015pi,borrmann2015extremely}. In these materials, linear bands cross near the Fermi level and form Dirac/Weyl points. The emergent quasi-particles are relativistic Dirac/Weyl fermions. Typical three dimensional (3D) Dirac semimetals Cd$_3$As$_2$\cite{neupane2014observationCd3As2,YLChen2014stableCd3As2,PhysRevLett.113.027603,liang2015ultrahigh,li2015giant,li2015negative}, Na$_3$Bi\cite{PhysRevB.85.195320,liu2014discovery,xiong2015evidence,xiong2016anomalous} and Weyl semimetals (Nb/Ta)(P/As)\cite{PhysRevX.5.011029,xu2015discoveryTaAs,huang2015weyl,PhysRevX.5.031013DingHongTaAs,NatPhysDingHongTaAs,yang2015weyl,xu2015discoveryNbAs,xu2015experimental,xu2016observation,liu2016evolution,PhysRevX.5.031023,zhang2016signatures,arnold2016negative,hu2015pi,borrmann2015extremely} have been researched deeply. Several exotic transport properties such as large linear magnetoresistance, high carrier mobility, nontrivial Berry phase, and chiral anomaly have been observed in them\cite{liang2015ultrahigh,xiong2016anomalous,PhysRevX.5.031023,borrmann2015extremely,hu2015pi,xiong2015evidence,zhang2016signatures,li2015giant,li2015negative}.

The Mn-based ternary 112 type compounds (Ca/Sr/Ba/Eu/Yb)MnBi$_2$\cite{PhysRevB.85.041101,he2012giant,feng2014strong,PhysRevB.87.245104,PhysRevLett.107.126402,PhysRevB.84.220401,PhysRevB.84.064428,PhysRevB.90.075120,PhysRevLett.113.156602,PhysRevB.90.035133,zhang2016interplay,Petrovic2016,wang2016large,masuda2016quantum,PhysRevB.90.075109,borisenko2015time,PhysRevB.94.165161} and (Ca/Sr/Ba/Yb)MnSb$_2$\cite{0953-8984-26-4-042201,liu2017discovery,liu2016nearly,PhysRevB.95.045128,kealhofer2017observation,wang2017Quantum} have been established as Dirac materials. Especially, YbMnBi$_2$\cite{borisenko2015time} and Sr$_{1-y}$Mn$_{1-z}$Sb$_2$\cite{liu2017discovery} have been further suggested to host the time reversal symmetry breaking Weyl fermions. The Bi/Sb square net in these compounds provides an ideal platform for Dirac/Weyl fermions. Since even single layer Bi square net can produce Dirac point\cite{PhysRevB.87.245104}, it is a feasible way to search for Dirac fermions in materials including such Bi square net.

Large unsaturated linear MR is an interesting transport phenomena in these Mn-based 112 type Dirac materials, in which the linear MR has been attributed to the extreme quantum limit of Dirac fermions\cite{PhysRevB.85.041101,PhysRevB.84.220401,Petrovic2016,wang2016large,PhysRevB.94.165161}. In fact, there exist various materials exhibiting linear MR, such as bismuth films\cite{yang1999large}, Ag$_{2+\delta}$Se/Ag$_{2+\delta}$Te\cite{xu1997large,PhysRevLett.88.066602,PhysRevLett.95.186603,PhysRevB.79.035204,PhysRevLett.106.156808}, \emph{n}-type narrow gap semiconductor InSb\cite{hu2008classical}, graphene\cite{friedman2010quantum,wang2014classical,kisslinger2015linear,PhysRevB.93.195430}, topological insulators Bi$_2$Se$_3$/Bi$_2$Te$_3$\cite{wang2014granularity,yan2013large,PhysRevLett.108.266806,tang2011two,he2013disorder}, and Dirac/Weyl semimetals\cite{liang2015ultrahigh,xiong2016anomalous,PhysRevX.5.031023,borrmann2015extremely,hu2015pi} \emph{etc}. Several theories have been developed to explain the linear MR. Abrikosov established the quantum linear MR theory for Dirac fermions in the quantum limit, where only the lowest Landau level is occupied\cite{PhysRevB.58.2788}. Parish and Littlewood provided a classical route (PL model) to linear MR, in which the inhomogeneity in disordered conductors is believed to produce linear MR\cite{parish2003non}. Hu \emph{et al.} expanded the PL model to 3D and identified conductivity fluctuations as the underlying physical mechanism\cite{PhysRevB.75.214203}. Kisslinger \emph{et al.} further pointed out that the linear MR in PL model is induced by charge carrier density fluctuations rather than mobility fluctuations\cite{PhysRevB.95.024204}. Recently, Alekseev \emph{et al.} proposed a new classical mechanism for linear MR in 2D two-component systems\cite{PhysRevLett.114.156601,PhysRevB.93.195430,alekseev2016magnetoresistance}. In the model, the lateral quasi-particle flow will lead to the accumulation of excess quasi-particles near the sample edges. It is believed that the interplay of bulk and edge resistances gives rise to the possibility of linear MR. The mechanism is not only suitable for charge neutrality (compensated) two-component systems, but also effective in the systems which stay away from charge neutrality.

Motivated by the possible emergence of Dirac fermion in BaZnBi$_2$, we synthesized the single crystals of BaZnBi$_2$ and studied the magneto-transport properties. The MR depends on the field linearly and angular-dependent magnetoresistance reveals the two-fold symmetry of Fermi surface in BaZnBi$_2$. SdH oscillation can be observed clearly at low temperature and high field. However, the Landau level index fan diagram yields a trivial Berry phase and the first-principles calculations indicate the absence of stable gapless Dirac fermion and the coexistence of electron-like and hole-like FSs in BaZnBi$_2$. Although the structure of BaZnBi$_2$ contains the single layer Bi square net, it is suggested to be a topologically trivial material by considering the trivial Berry phase and band structures from first-principles calculations.

\section{Experimental methods and crystal structure}

Single crystals of BaZnBi$_2$ were grown from Bi flux. Ba, Zn and excess Bi were placed in a crucible and sealed in a quartz tube with a ratio of Ba:Zn:Bi=1:1:6. The quartz tube was heated to 1000$^0$C, held there for 15 h, and cooled to 370$^0$C at a rate of 2$^0$C/h, then the excess Bi-flux was removed by centrifugation. Once the samples were cooled to room temperature, the silver, plate-shaped crystals were obtained. The crystals can be cleaved easily. Elemental analysis was performed using energy dispersive x-ray spectroscopy (EDS, Oxford X-Max 50). The determined atomic proportion was consistent with the composition of BaZnBi$_2$ within instrumental error. Single crystal and powder x-ray diffraction (XRD) patterns were collected from a Bruker D8 Advance x-ray diffractometer using Cu K$_{\alpha}$ radiation. TOPAS-4.2 was employed for the refinement. Resistivity measurements were performed on a Quantum Design physical property measurement system (QD PPMS-14T). The electronic structures of BaZnBi$_2$ have been studied by using first-principles calculations. The projector augmented wave (PAW) method\cite{PhysRevB.50.17953,PhysRevB.59.1758} as implemented in the VASP package\cite{PhysRevB.47.558,kresse1996efficiency,PhysRevB.54.11169} was used to describe the core electrons. For the exchange-correlation potential, the generalized gradient approximation (GGA) of Perdew-Burke-Ernzerhof formula\cite{PhysRevLett.77.3865} was adopted. The kinetic energy cutoff of the plane-wave basis was set to be 360 eV. A 20$\times$20$\times$20 \emph{k}-point mesh was utilized for the Brillouin zone (BZ) sampling and the Fermi surface was broadened by a Gaussian smearing method with a width of 0.05 eV. The electronic structures were calculated both without and with the spin-orbital coupling (SOC) effect. The Fermi surfaces were studied by using the maximally localized Wannier functions (MLWF)\cite{PhysRevB.56.12847,PhysRevB.65.035109} method and the carrier concentrations were analyzed based on the information of Fermi surface volumes. As shown in Fig. 1(a), BaZnBi$_2$ is isostructural with (Sr/Ba/Eu)MnBi$_2$. BaZnBi$_2$ is comprised of alternating ZnBi and BaBi layers\cite{brechtel1981neue}. In the ZnBi layer, each Zn atom is surrounded by four Bi atoms, which form the ZnBi$_4$ tetrahedra. In the BaBi layer, Bi atoms are separated and form a square net. Figures 2(a) and 2(b) show the powder and single crystal x-ray diffraction patterns of BaZnBi$_2$, respectively. The powder diffraction pattern was refined by TOPAS and could be well indexed in the $I4/mmm$ space group. The determined lattice parameters are $a=b=4.853(8){\AA}$ and $c=22.011(9){\AA}$, in agreement with the previous published data\cite{brechtel1981neue}. The single crystal XRD pattern indicates that the plane is normal to the \emph{c} axis of the crystal.

\begin{figure}[htbp]
\centering
\includegraphics[width=0.48\textwidth]{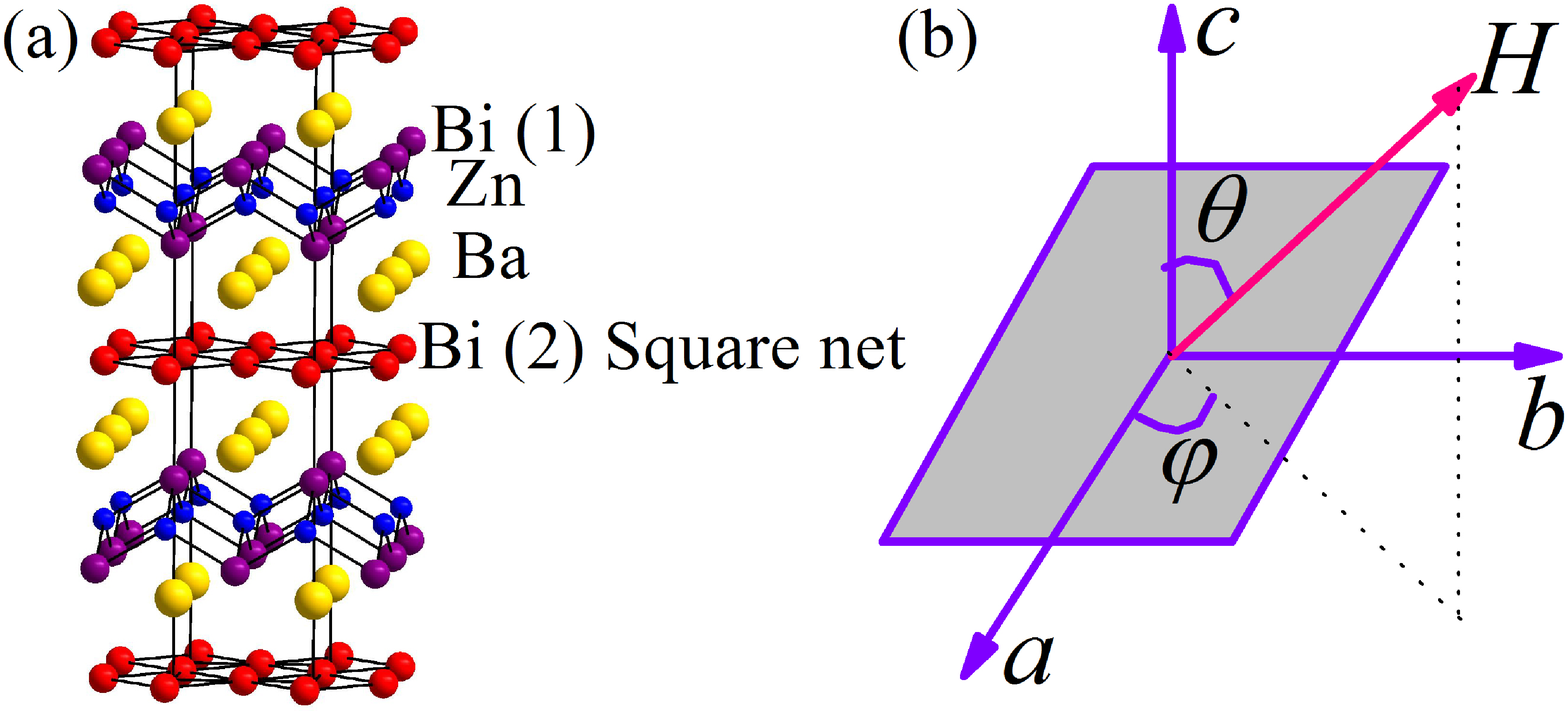}
\caption{(a) The crystal structure of BaZnBi$_2$. (b) The definition of polar angle $\theta$ and azimuthal angle $\varphi$.}
\end{figure}

\begin{figure}[htbp]
\centering
\includegraphics[width=0.48\textwidth]{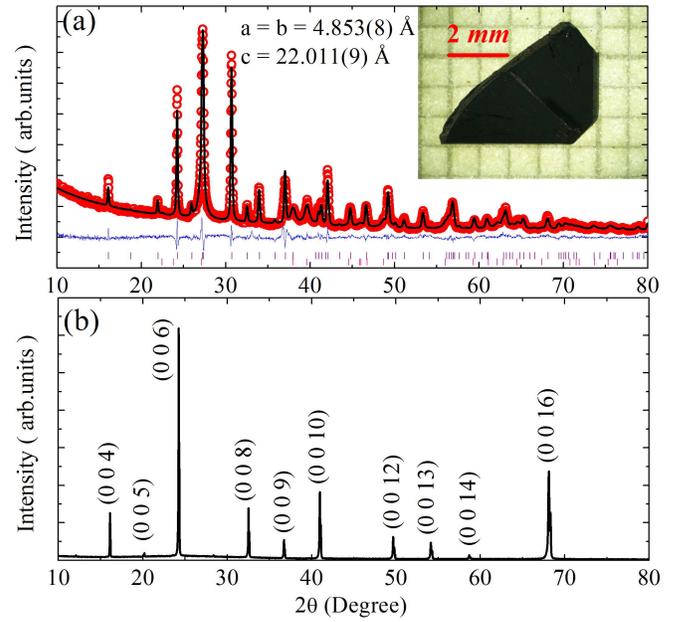}
\caption{(a) Refinement results of powder samples using TOPAS. Red circle indicates the data of experiment. The fit is given by black solid line. The difference plot is in blue. The purple and pink vertical lines denote the positions of Bragg peaks of BaZnBi$_2$ and Bi, respectively. The inset is an image of a typical single crystal. (b)Single crystal x-ray diffraction pattern of a BaZnBi$_2$ crystal, showing only the (\emph{00l}) reflections.}
\end{figure}

\section{Results and discussions}

As shown in Fig. 3(a), the in-plane resistivity $\rho_{ab}(T)$ exhibits metallic behavior, and an external magnetic field can slightly enhance the resistivity. The MR nearly disappears when the temperature increases above 100 K. Figure 3(b) shows the magnetic field dependence of in-plane MR at different temperatures. According to classical theory, the MR follows a quadratic \emph{H} dependence and reaches saturation at high field. However, in BaZnBi$_2$, MR increases linearly and has no sign of saturation up to 14 T. The critical point $H_c$ of the crossover from quadratic to linear dependence on \emph{H} is very small (at 2.2 K, $H_c$$\approx$0.43 T, not presented here). The origin of the linear MR will be discussed below. The Kohler plot of BaZnBi$_2$ is displayed in the inset of Fig. 3(b). According to the Kohler's rule $MR=F(H/\rho_0)$ ($\rho_0$ is the resistivity under zero field), MR measured at different temperatures can be expressed solely as a function of $H/\rho_0$ if there is a single scattering time in the material. It is clear that the curves do not fall into one curve. The violation of Kohler's rule indicates that there is more than one type of carrier in BaZnBi$_2$\cite{PhysRevB.57.11854}, which has also been confirmed by our first-principles calculations.

\begin{figure}[htbp]
\centering
\includegraphics[width=0.48\textwidth]{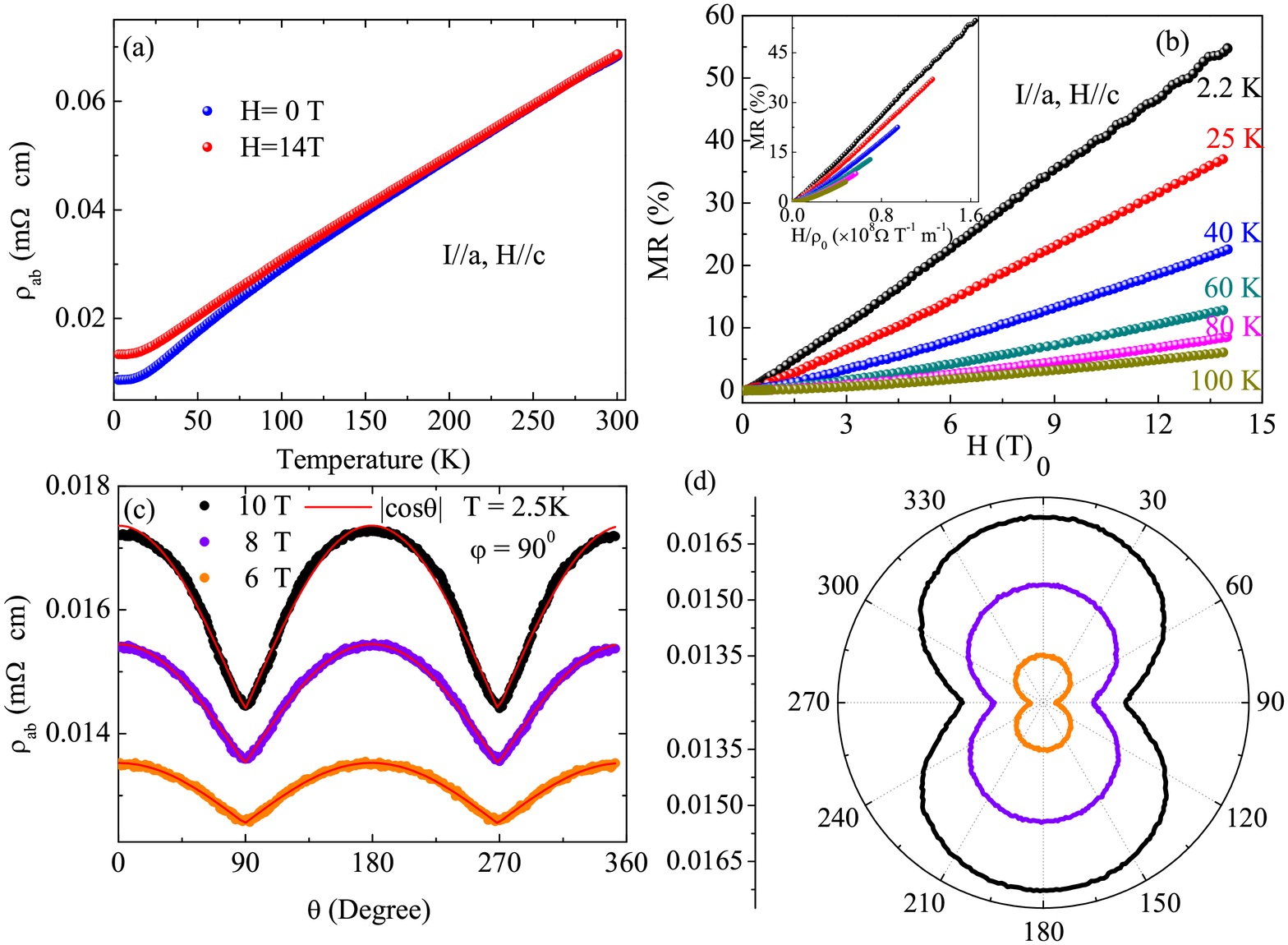}
\caption{(a) Temperature dependence of in-plane resistivity $\rho_{ab}(T)$ under \emph{H}=0 T and 14 T. (b) In-plane MR vs magnetic field \emph{H} at different temperatures. Inset: Kohler plot of BaZnBi$_2$ at different temperatures. (c) Polar angle $\theta$ dependence of in-plane resistivity $\rho_{ab}(\theta)$ at \emph{H}=10,8,6 T. The red curve is the fitting of the function $|cos\theta|$. The temperature and azimuthal angle are fixed at \emph{T}=2.5 K and $\varphi=90^0$ respectively. (d) The corresponding polar plot for $\rho_{ab}(\theta)$.}
\end{figure}

When a metal is in the magnetic field, the Lorentz force affects the momentum components of the carriers in the plane perpendicular to the field, consequently, the MR is partially determined by the mobility in this plane. For a quasi-2D material, the carriers will only respond to the magnetic field component $H|cos\theta|$. Figure 3(c) shows the angular-dependent in-plane resistivity $\rho_{ab}(\theta)$ under different magnetic fields ($\varphi$ was fixed at $90^0$). When the magnetic field is parallel to \emph{c} axis ($\theta=0^0,180^0,360^0$), resistivity has the maximum value. With the increase in the polar angle $\theta$, the resistivity decreases gradually and reaches the minimum value when the magnetic field is parallel to \emph{ab} plane ($\theta=90^0,270^0$). The curves follow the function of $|cos\theta|$ very well as fitted in the figure (red lines). The behavior of angular-dependent in-plane resistivity suggests the dominance of quasi-2D Fermi surface in BaZnBi$_2$. Figure 3(d) shows the polar plot for $\rho_{ab}(\theta)$, which exhibits the two-fold symmetry of Fermi surface clearly.

\begin{figure}[htbp]
\centering
\includegraphics[width=0.48\textwidth]{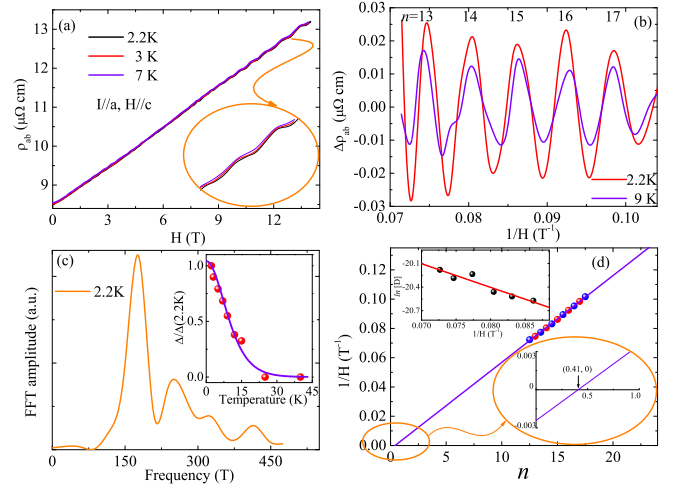}
\caption{(a) Magnetic field dependence of resistivity of BaZnBi$_2$ at low temperatures. The inset shows the enlarged part of $\rho_{ab}(H)$ at high field, where the oscillation can be observed more clearly. (b) The amplitude of SdH oscillations plotted as a function of the reciprocal of magnetic field. (c) The FFT spectrum of the corresponding SdH oscillation at 2.2 K for BaZnBi$_2$. Inset: Temperature dependence of the relative FFT amplitude of oscillation frequency. The violet solid line is the fitting result based on Lifshitz-Kosevich formula, which yields the effective cyclotron resonant mass. (d) Landau index \emph{n} plotted against $1/H$. The peaks (red circles) and valleys (blue circles) of the $\Delta \rho_{ab}$ at 2.2 K are assigned as integer and half integer LL indices, respectively. Upper inset: \emph{ln}[D] (defined in the text) plotted as a function of $1/H$. The Dingle temperature $T_D$ is calculated from the slope of the linear fit. Lower inset: Enlargement of the intercept of linear fit.}
\end{figure}

\begin{figure}[htbp]
\centering
\includegraphics[width=0.48\textwidth]{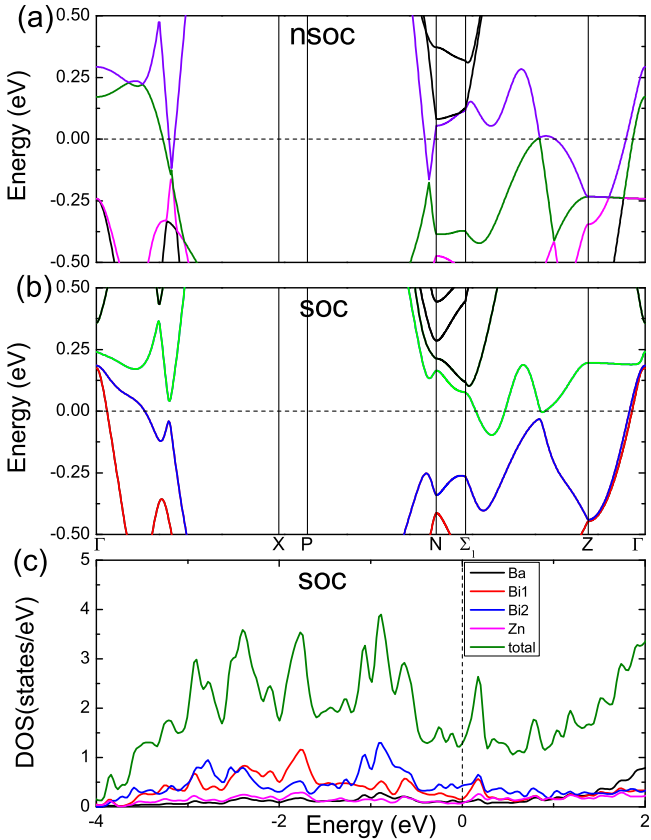}
\caption{The band structures of BaZnBi$_2$ calculated (a) without and (b) with the spin-orbital coupling (SOC) effect. (c) Total and local density of states calculated with the SOC effect.}
\end{figure}

Figure 4(a) shows the resistivity of BaZnBi$_2$ as a function of field. SdH oscillation was observed at low temperature and high field. Figure 4(b) plots the oscillation amplitude $\Delta\rho_{ab}=\rho_{ab}-\langle\rho_{ab}\rangle$ against the reciprocal of magnetic field $1/H$. A main oscillation frequency \emph{F}=178 T is identified from the fast Fourier transformation (FFT) spectra (Fig. 4(c)). Another weak frequency 251 T can also be seen from the spectra, which may be contributed by the hole-like FSs. We analyze the principal frequency 178 T as follows. According to the Onsager relation $F=(\phi_0/2\pi^2)A=(\hbar/2\pi e)A$, the cross sectional area \emph{A} of Fermi surface normal to the magnetic field is $1.7\times 10^{-2}{\AA}^{-2}$. The Fermi wave vector is $k_F=0.07{\AA}^{-1}$ by assuming a circular cross section. The amplitude of SdH oscillation can be described using Lifshitz-Kosevich formula,
\begin{equation}\label{equ1}
\Delta\rho\propto\frac{\lambda T}{sinh(\lambda T)}e^{-\lambda T_D}cos[2\pi\times(\frac{F}{H}-\frac{1}{2}+\beta+\delta)]
\end{equation}
where $\lambda= (2\pi^2k_{B}m^*)/(\hbar e\bar{H})$ and $T_D$ is Dingle temperature. $2\pi \beta$ is the Berry phase. $\delta$ is a phase shift with the value of $\delta=0$ (or $\pm1/8$) for 2D (or 3D) system. The temperature dependence of relative FFT amplitude can be well fitted based on the thermal factor $R_T=(\lambda T)/sinh(\lambda T)$ in L-K formula as shown in the inset of Fig. 4(c). The cyclotron effective mass yielded by the fitting is $m^*=0.18m_e$. According to the relations $v_F=\hbar k_F/m^*$ and $m^*=E_F/v^2_F$, the Fermi velocity $v_F=4.7\times 10^5m/s$ and Fermi energy $E_F=0.226eV$ can also be obtained. Figure 4(d) shows $1/H$ as a function of Landau index \emph{n}. The Landau index \emph{n} is related to $1/H$ by Lifshitz-Onsager quantization rule $A(\hbar/2\pi eH)=n+1/2-\beta-\delta$. Considering the dominant 2D Fermi surface in BaZnBi$_2$, the value of $\delta$ is close to zero. So the intercept ( the lower inset in Fig. 4(d)) of linear fit gives a Berry phase $2\pi\beta=1.82\pi$. As is well known, the Berry phase is usually 0 or $2\pi$ in normal metals with trivial parabolic dispersion relation, and the non-trivial $\pi$ Berry phase is a transport characteristic for materials with gapless Dirac linear dispersion. The value of $2\pi\beta$ in BaZnBi$_2$ is far away from the non-trivial value $\pi$, indicating BaZnBi$_2$ is possibly a normal metal. The upper inset in Fig. 4(d) shows the semilog plot of $D=\Delta\rho Hsinh(\lambda T)$ versus $1/H$ at 2.2 K. The fitted Dingle temperature $T_D$ is 11.2 K, corresponding to a carrier lifetime $\tau_q=\hbar/(2\pi k_B T_D)=1.08(5)\times 10^{-13}s$ and quantum mobility $\mu_q=e\tau_q/m^*=1047cm^2/Vs$.

\begin{figure*}[htbp]
\centering
\includegraphics[width=\textwidth]{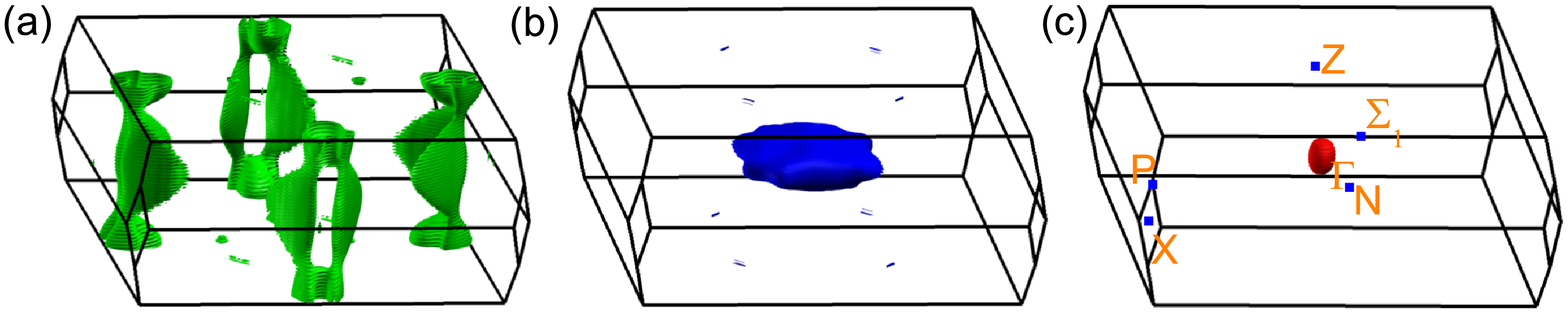}
\caption{The (a) electron-like and (b)(c) hole-like Fermi surface sheets of BaZnBi$_2$. The colors of Fermi surface sheets and the bands crossing the Fermi level in Fig. 5(b) are in one-to-one correspondence.}
\end{figure*}

Considering that the magnetic field is not high enough to drive the system into its quantum limit, the band structures of BaZnBi$_2$ have been studied by using first-principles calculations. Without the spin-orbital coupling (SOC) effect, BaZnBi$_2$ shows many Dirac points around the Fermi level (Fig. 5(a)) along the high-symmetry  $\Gamma$-X, P-N, M-Z, and  $\Gamma$-Z paths in Brillouin zone (BZ) (Fig. 6(c)). Considering the $C_{4v}$ double group symmetry of BaZnBi$_2$, only the Dirac points along the  $\Gamma$-Z path would be protected when including the SOC effect, while others would open gaps. Nevertheless, once the SOC effect is included, the band structures of BaZnBi$_2$ around the Fermi level show dramatic changes (Fig. 5(b)), and even the Dirac point along the  $\Gamma$-Z path at 0.25 eV below Fermi level disappears. This is because the states around Fermi level are mainly contributed by Bi2 atoms (Fig. 1(a)), as deduced from the local density of states (Fig. 5(c)), and the strong SOC effect in heavy element Bi has a great influence on related bands.

On the other hand, as shown in Fig. 5(b), the valence bands and conducting bands of BaZnBi$_2$ have little overlaps. The corresponding FSs of these bands are displayed in Fig. 6. Both the electron-like (Fig. 6(a)) FSs around N-M path and one of the hole-like (Fig. 6(b)) FSs around $\Gamma$ point possess 2D characteristics, while the other hole-like FS around $\Gamma$ point exhibits a small ellipsoidal shape. By calculating the volumes of these FSs, we obtain the electron-like and hole-like carrier concentrations as $n_e=1.88\times10^{20} cm^{-3}$ and $n_h=1.79\times10^{20} cm^{-3}$, respectively. This indicates that BaZnBi$_2$ is a charge-compensated semimetal\cite{PhysRevB.93.235142}.

The linear MR in BaZnBi$_2$ is an interesting phenomenon. According to the theory of quantum MR proposed by Abrikosov\cite{PhysRevB.58.2788}, linear MR will emerge when a system with gapless linear band dispersion enters the quantum limit, where all the carriers occupy the lowest Landau level. Quantum linear MR has been achieved in \emph{n}-type InSb\cite{hu2008classical} and Bi$_2$Te$_3$ nanosheets\cite{PhysRevLett.108.266806}. The linear MR in Mn-based 112 Dirac materials (Ca/Sr/Ba/Eu/Yb)MnBi$_2$ is also suggested to be quantum MR\cite{PhysRevB.85.041101,PhysRevB.84.220401,Petrovic2016,wang2016large,PhysRevB.90.075109,PhysRevB.94.165161}. However, the quantum MR is excluded in BaZnBi$_2$ for the following reasons: (i) The first-principles calculations reveal the absence of stable gapless Dirac cone in BaZnBi$_2$. Considering the trivial Berry phase extracted from the Landau level index fan diagram, BaZnBi$_2$ is possibly not a Dirac material. (ii) The condition for quantum limit is $n\ll(eH/\hbar)^{3/2}$, where $n$ is the carrier concentration. Substituting $n_e(n_h)$ into the expression, the quantum limit can only be reached when $H\gg(\hbar/e)n^{2/3}\approx$ 216 T(209 T), which is not possible in the lab. Although the linear MR may emerge at lower field with more than the lowest Landau level filled\cite{hu2008classical}, the onset magnetic field of linear MR in BaZnBi$_2$ is too far from the quantum limit.

The classical PL model\cite{parish2003non,kozlova2012linear} is usually used to explain the linear MR. In a 2D disordered conductor, mobility fluctuations (later modified as charge carrier density fluctuations by Kisslinger \emph{et al.}\cite{PhysRevB.95.024204}) induce the distortion of current path, the mixture of the transverse Hall resistivity gives rise to a linear MR. The model has been expanded to the 3D case\cite{PhysRevB.75.214203}. Hall measurements are usually performed to confirm the role of disorder in the generation of linear MR. However, the tests on Hall resistivity are not successful. It's difficult to obtain a smooth curve of the field dependent Hall resistivity, which may be due to the small Hall resistivity caused by the nearly compensated carrier concentrations. As the samples' residual resistivity ratio is small (RRR$\approx$8), disorder is inevitable and may contribute to the linear MR.

Another classical mechanism was presented by Alekseev \emph{et al.} for the linear MR in 2D two-component systems\cite{PhysRevLett.114.156601,PhysRevB.93.195430,alekseev2016magnetoresistance}. According to the model, electron-hole recombination allows for a lateral quasi-particle flow and excess quasi-particles will accumulate near the edges of sample, the rise of edge conductivity will lead to the emergence of linear MR. At the charge neutrality point, the total resistivity $\rho_{total}$ can be expressed as\cite{PhysRevLett.114.156601},
\begin{equation}\label{equ2}
\rho_{total}^{-1}=eP\mu [\frac{1}{(\mu H)^2}+\frac{l_0}{W\mu H}]
\end{equation}
where $P=n_e+n_h$ is the quasiparticle density, $l_0$ is the recombination length at zero field, $W$ is the width of sample, and $\mu$ is the mobility of carriers. Thus, the second term in Eq. (2) will be far beyond the first term in high field, so the MR exhibits linear behavior. As the electron and hole are nearly compensated ($n_h/n_e\approx0.95$) in BaZnBi$_2$, this theory is also possible in explaining the observed linear MR. Further experiments are needed to clarify the contribution of disorder and edge conductivity to the linear MR in BaZnBi$_2$.

\section{Summary}

In summary, single crystals of BaZnBi$_2$ have been grown. BaZnBi$_2$ exhibits two-fold symmetry magneto-transport properties and linear MR. SdH oscillations are observed at low temperature and high field, and the extracted Berry phase from the Landau level index fan diagram is topologically trivial. In the first-principles calculations, no gapless Dirac point is found when the SOC effect is included. The obtained carrier concentrations indicate that BaZnBi$_2$ is a charge-compensated semimetal. The origin of linear MR is also discussed.

\section{Acknowledgments}

We thank Shuyun Zhou and Zhong-Yi Lu for helpful discussions. This work is supported by the National Natural Science Foundation of China (No.11574391), the Fundamental Research Funds for the Central Universities, and the Research Funds of Renmin University of China (No.14XNLQ03 and 14XNLQ07). Computational resources have been provided by the Physical Laboratory of High Performance Computing at RUC. The Fermi surfaces were prepared with the XCRYSDEN program\cite{kokalj2003computer}.

\bibliography{bibtex}

\begin{thebibliography}{86}%
\makeatletter
\providecommand \@ifxundefined [1]{%
 \@ifx{#1\undefined}
}%
\providecommand \@ifnum [1]{%
 \ifnum #1\expandafter \@firstoftwo
 \else \expandafter \@secondoftwo
 \fi
}%
\providecommand \@ifx [1]{%
 \ifx #1\expandafter \@firstoftwo
 \else \expandafter \@secondoftwo
 \fi
}%
\providecommand \natexlab [1]{#1}%
\providecommand \enquote  [1]{``#1''}%
\providecommand \bibnamefont  [1]{#1}%
\providecommand \bibfnamefont [1]{#1}%
\providecommand \citenamefont [1]{#1}%
\providecommand \href@noop [0]{\@secondoftwo}%
\providecommand \href [0]{\begingroup \@sanitize@url \@href}%
\providecommand \@href[1]{\@@startlink{#1}\@@href}%
\providecommand \@@href[1]{\endgroup#1\@@endlink}%
\providecommand \@sanitize@url [0]{\catcode `\\12\catcode `\$12\catcode
  `\&12\catcode `\#12\catcode `\^12\catcode `\_12\catcode `\%12\relax}%
\providecommand \@@startlink[1]{}%
\providecommand \@@endlink[0]{}%
\providecommand \url  [0]{\begingroup\@sanitize@url \@url }%
\providecommand \@url [1]{\endgroup\@href {#1}{\urlprefix }}%
\providecommand \urlprefix  [0]{URL }%
\providecommand \Eprint [0]{\href }%
\providecommand \doibase [0]{http://dx.doi.org/}%
\providecommand \selectlanguage [0]{\@gobble}%
\providecommand \bibinfo  [0]{\@secondoftwo}%
\providecommand \bibfield  [0]{\@secondoftwo}%
\providecommand \translation [1]{[#1]}%
\providecommand \BibitemOpen [0]{}%
\providecommand \bibitemStop [0]{}%
\providecommand \bibitemNoStop [0]{.\EOS\space}%
\providecommand \EOS [0]{\spacefactor3000\relax}%
\providecommand \BibitemShut  [1]{\csname bibitem#1\endcsname}%
\let\auto@bib@innerbib\@empty
\bibitem [{\citenamefont {Wan}\ \emph {et~al.}(2011)\citenamefont {Wan},
  \citenamefont {Turner}, \citenamefont {Vishwanath},\ and\ \citenamefont
  {Savrasov}}]{PhysRevB.83.205101}%
  \BibitemOpen
  \bibfield  {author} {\bibinfo {author} {\bibfnamefont {X.}~\bibnamefont
  {Wan}}, \bibinfo {author} {\bibfnamefont {A.~M.}\ \bibnamefont {Turner}},
  \bibinfo {author} {\bibfnamefont {A.}~\bibnamefont {Vishwanath}}, \ and\
  \bibinfo {author} {\bibfnamefont {S.~Y.}\ \bibnamefont {Savrasov}},\ }\href
  {\doibase 10.1103/PhysRevB.83.205101} {\bibfield  {journal} {\bibinfo
  {journal} {Phys. Rev. B}\ }\textbf {\bibinfo {volume} {83}},\ \bibinfo
  {pages} {205101} (\bibinfo {year} {2011})}\BibitemShut {NoStop}%
\bibitem [{\citenamefont {Wehling}\ \emph {et~al.}(2014)\citenamefont
  {Wehling}, \citenamefont {Black-Schaffer},\ and\ \citenamefont
  {Balatsky}}]{wehling2014dirac}%
  \BibitemOpen
  \bibfield  {author} {\bibinfo {author} {\bibfnamefont {T.}~\bibnamefont
  {Wehling}}, \bibinfo {author} {\bibfnamefont {A.~M.}\ \bibnamefont
  {Black-Schaffer}}, \ and\ \bibinfo {author} {\bibfnamefont {A.~V.}\
  \bibnamefont {Balatsky}},\ }\href@noop {} {\bibfield  {journal} {\bibinfo
  {journal} {Adv. Phys.}\ }\textbf {\bibinfo {volume} {63}},\ \bibinfo {pages}
  {1} (\bibinfo {year} {2014})}\BibitemShut {NoStop}%
\bibitem [{\citenamefont {Wang}\ \emph
  {et~al.}(2012{\natexlab{a}})\citenamefont {Wang}, \citenamefont {Sun},
  \citenamefont {Chen}, \citenamefont {Franchini}, \citenamefont {Xu},
  \citenamefont {Weng}, \citenamefont {Dai},\ and\ \citenamefont
  {Fang}}]{PhysRevB.85.195320}%
  \BibitemOpen
  \bibfield  {author} {\bibinfo {author} {\bibfnamefont {Z.}~\bibnamefont
  {Wang}}, \bibinfo {author} {\bibfnamefont {Y.}~\bibnamefont {Sun}}, \bibinfo
  {author} {\bibfnamefont {X.-Q.}\ \bibnamefont {Chen}}, \bibinfo {author}
  {\bibfnamefont {C.}~\bibnamefont {Franchini}}, \bibinfo {author}
  {\bibfnamefont {G.}~\bibnamefont {Xu}}, \bibinfo {author} {\bibfnamefont
  {H.}~\bibnamefont {Weng}}, \bibinfo {author} {\bibfnamefont {X.}~\bibnamefont
  {Dai}}, \ and\ \bibinfo {author} {\bibfnamefont {Z.}~\bibnamefont {Fang}},\
  }\href {\doibase 10.1103/PhysRevB.85.195320} {\bibfield  {journal} {\bibinfo
  {journal} {Phys. Rev. B}\ }\textbf {\bibinfo {volume} {85}},\ \bibinfo
  {pages} {195320} (\bibinfo {year} {2012}{\natexlab{a}})}\BibitemShut
  {NoStop}%
\bibitem [{\citenamefont {Weng}\ \emph {et~al.}(2016)\citenamefont {Weng},
  \citenamefont {Dai},\ and\ \citenamefont {Fang}}]{weng2016topological}%
  \BibitemOpen
  \bibfield  {author} {\bibinfo {author} {\bibfnamefont {H.}~\bibnamefont
  {Weng}}, \bibinfo {author} {\bibfnamefont {X.}~\bibnamefont {Dai}}, \ and\
  \bibinfo {author} {\bibfnamefont {Z.}~\bibnamefont {Fang}},\ }\href@noop {}
  {\bibfield  {journal} {\bibinfo  {journal} {J. Phys.-Condes. Matter}\
  }\textbf {\bibinfo {volume} {28}},\ \bibinfo {pages} {303001} (\bibinfo
  {year} {2016})}\BibitemShut {NoStop}%
\bibitem [{\citenamefont {Liu}\ \emph {et~al.}(2014{\natexlab{a}})\citenamefont
  {Liu}, \citenamefont {Zhou}, \citenamefont {Zhang}, \citenamefont {Wang},
  \citenamefont {Weng}, \citenamefont {Prabhakaran}, \citenamefont {Mo},
  \citenamefont {Shen}, \citenamefont {Fang}, \citenamefont {Dai} \emph
  {et~al.}}]{liu2014discovery}%
  \BibitemOpen
  \bibfield  {author} {\bibinfo {author} {\bibfnamefont {Z.}~\bibnamefont
  {Liu}}, \bibinfo {author} {\bibfnamefont {B.}~\bibnamefont {Zhou}}, \bibinfo
  {author} {\bibfnamefont {Y.}~\bibnamefont {Zhang}}, \bibinfo {author}
  {\bibfnamefont {Z.}~\bibnamefont {Wang}}, \bibinfo {author} {\bibfnamefont
  {H.}~\bibnamefont {Weng}}, \bibinfo {author} {\bibfnamefont {D.}~\bibnamefont
  {Prabhakaran}}, \bibinfo {author} {\bibfnamefont {S.-K.}\ \bibnamefont {Mo}},
  \bibinfo {author} {\bibfnamefont {Z.}~\bibnamefont {Shen}}, \bibinfo {author}
  {\bibfnamefont {Z.}~\bibnamefont {Fang}}, \bibinfo {author} {\bibfnamefont
  {X.}~\bibnamefont {Dai}},  \emph {et~al.},\ }\href@noop {} {\bibfield
  {journal} {\bibinfo  {journal} {Science}\ }\textbf {\bibinfo {volume}
  {343}},\ \bibinfo {pages} {864} (\bibinfo {year}
  {2014}{\natexlab{a}})}\BibitemShut {NoStop}%
\bibitem [{\citenamefont {Neupane}\ \emph {et~al.}(2014)\citenamefont
  {Neupane}, \citenamefont {Xu}, \citenamefont {Sankar}, \citenamefont
  {Alidoust}, \citenamefont {Bian}, \citenamefont {Liu}, \citenamefont
  {Belopolski}, \citenamefont {Chang}, \citenamefont {Jeng}, \citenamefont
  {Lin} \emph {et~al.}}]{neupane2014observationCd3As2}%
  \BibitemOpen
  \bibfield  {author} {\bibinfo {author} {\bibfnamefont {M.}~\bibnamefont
  {Neupane}}, \bibinfo {author} {\bibfnamefont {S.-Y.}\ \bibnamefont {Xu}},
  \bibinfo {author} {\bibfnamefont {R.}~\bibnamefont {Sankar}}, \bibinfo
  {author} {\bibfnamefont {N.}~\bibnamefont {Alidoust}}, \bibinfo {author}
  {\bibfnamefont {G.}~\bibnamefont {Bian}}, \bibinfo {author} {\bibfnamefont
  {C.}~\bibnamefont {Liu}}, \bibinfo {author} {\bibfnamefont {I.}~\bibnamefont
  {Belopolski}}, \bibinfo {author} {\bibfnamefont {T.-R.}\ \bibnamefont
  {Chang}}, \bibinfo {author} {\bibfnamefont {H.-T.}\ \bibnamefont {Jeng}},
  \bibinfo {author} {\bibfnamefont {H.}~\bibnamefont {Lin}},  \emph {et~al.},\
  }\href@noop {} {\bibfield  {journal} {\bibinfo  {journal} {Nat. Commun.}\
  }\textbf {\bibinfo {volume} {5}},\ \bibinfo {pages} {3786} (\bibinfo {year}
  {2014})}\BibitemShut {NoStop}%
\bibitem [{\citenamefont {Liu}\ \emph {et~al.}(2014{\natexlab{b}})\citenamefont
  {Liu}, \citenamefont {Jiang}, \citenamefont {Zhou}, \citenamefont {Wang},
  \citenamefont {Zhang}, \citenamefont {Weng}, \citenamefont {Prabhakaran},
  \citenamefont {Mo}, \citenamefont {Peng}, \citenamefont {Dudin} \emph
  {et~al.}}]{YLChen2014stableCd3As2}%
  \BibitemOpen
  \bibfield  {author} {\bibinfo {author} {\bibfnamefont {Z.}~\bibnamefont
  {Liu}}, \bibinfo {author} {\bibfnamefont {J.}~\bibnamefont {Jiang}}, \bibinfo
  {author} {\bibfnamefont {B.}~\bibnamefont {Zhou}}, \bibinfo {author}
  {\bibfnamefont {Z.}~\bibnamefont {Wang}}, \bibinfo {author} {\bibfnamefont
  {Y.}~\bibnamefont {Zhang}}, \bibinfo {author} {\bibfnamefont
  {H.}~\bibnamefont {Weng}}, \bibinfo {author} {\bibfnamefont {D.}~\bibnamefont
  {Prabhakaran}}, \bibinfo {author} {\bibfnamefont {S.}~\bibnamefont {Mo}},
  \bibinfo {author} {\bibfnamefont {H.}~\bibnamefont {Peng}}, \bibinfo {author}
  {\bibfnamefont {P.}~\bibnamefont {Dudin}},  \emph {et~al.},\ }\href@noop {}
  {\bibfield  {journal} {\bibinfo  {journal} {Nat. Mater.}\ }\textbf {\bibinfo
  {volume} {13}},\ \bibinfo {pages} {677} (\bibinfo {year}
  {2014}{\natexlab{b}})}\BibitemShut {NoStop}%
\bibitem [{\citenamefont {Borisenko}\ \emph {et~al.}(2014)\citenamefont
  {Borisenko}, \citenamefont {Gibson}, \citenamefont {Evtushinsky},
  \citenamefont {Zabolotnyy}, \citenamefont {B\"uchner},\ and\ \citenamefont
  {Cava}}]{PhysRevLett.113.027603}%
  \BibitemOpen
  \bibfield  {author} {\bibinfo {author} {\bibfnamefont {S.}~\bibnamefont
  {Borisenko}}, \bibinfo {author} {\bibfnamefont {Q.}~\bibnamefont {Gibson}},
  \bibinfo {author} {\bibfnamefont {D.}~\bibnamefont {Evtushinsky}}, \bibinfo
  {author} {\bibfnamefont {V.}~\bibnamefont {Zabolotnyy}}, \bibinfo {author}
  {\bibfnamefont {B.}~\bibnamefont {B\"uchner}}, \ and\ \bibinfo {author}
  {\bibfnamefont {R.~J.}\ \bibnamefont {Cava}},\ }\href {\doibase
  10.1103/PhysRevLett.113.027603} {\bibfield  {journal} {\bibinfo  {journal}
  {Phys. Rev. Lett.}\ }\textbf {\bibinfo {volume} {113}},\ \bibinfo {pages}
  {027603} (\bibinfo {year} {2014})}\BibitemShut {NoStop}%
\bibitem [{\citenamefont {Liang}\ \emph {et~al.}(2015)\citenamefont {Liang},
  \citenamefont {Gibson}, \citenamefont {Ali}, \citenamefont {Liu},
  \citenamefont {Cava},\ and\ \citenamefont {Ong}}]{liang2015ultrahigh}%
  \BibitemOpen
  \bibfield  {author} {\bibinfo {author} {\bibfnamefont {T.}~\bibnamefont
  {Liang}}, \bibinfo {author} {\bibfnamefont {Q.}~\bibnamefont {Gibson}},
  \bibinfo {author} {\bibfnamefont {M.~N.}\ \bibnamefont {Ali}}, \bibinfo
  {author} {\bibfnamefont {M.}~\bibnamefont {Liu}}, \bibinfo {author}
  {\bibfnamefont {R.}~\bibnamefont {Cava}}, \ and\ \bibinfo {author}
  {\bibfnamefont {N.}~\bibnamefont {Ong}},\ }\href@noop {} {\bibfield
  {journal} {\bibinfo  {journal} {Nat. Mater.}\ }\textbf {\bibinfo {volume}
  {14}},\ \bibinfo {pages} {280} (\bibinfo {year} {2015})}\BibitemShut
  {NoStop}%
\bibitem [{\citenamefont {Weng}\ \emph {et~al.}(2015)\citenamefont {Weng},
  \citenamefont {Fang}, \citenamefont {Fang}, \citenamefont {Bernevig},\ and\
  \citenamefont {Dai}}]{PhysRevX.5.011029}%
  \BibitemOpen
  \bibfield  {author} {\bibinfo {author} {\bibfnamefont {H.}~\bibnamefont
  {Weng}}, \bibinfo {author} {\bibfnamefont {C.}~\bibnamefont {Fang}}, \bibinfo
  {author} {\bibfnamefont {Z.}~\bibnamefont {Fang}}, \bibinfo {author}
  {\bibfnamefont {B.~A.}\ \bibnamefont {Bernevig}}, \ and\ \bibinfo {author}
  {\bibfnamefont {X.}~\bibnamefont {Dai}},\ }\href@noop {} {\bibfield
  {journal} {\bibinfo  {journal} {Phys. Rev. X}\ }\textbf {\bibinfo {volume}
  {5}},\ \bibinfo {pages} {011029} (\bibinfo {year} {2015})}\BibitemShut
  {NoStop}%
\bibitem [{\citenamefont {Xu}\ \emph {et~al.}(2015{\natexlab{a}})\citenamefont
  {Xu}, \citenamefont {Belopolski}, \citenamefont {Alidoust}, \citenamefont
  {Neupane}, \citenamefont {Bian}, \citenamefont {Zhang}, \citenamefont
  {Sankar}, \citenamefont {Chang}, \citenamefont {Yuan}, \citenamefont {Lee}
  \emph {et~al.}}]{xu2015discoveryTaAs}%
  \BibitemOpen
  \bibfield  {author} {\bibinfo {author} {\bibfnamefont {S.-Y.}\ \bibnamefont
  {Xu}}, \bibinfo {author} {\bibfnamefont {I.}~\bibnamefont {Belopolski}},
  \bibinfo {author} {\bibfnamefont {N.}~\bibnamefont {Alidoust}}, \bibinfo
  {author} {\bibfnamefont {M.}~\bibnamefont {Neupane}}, \bibinfo {author}
  {\bibfnamefont {G.}~\bibnamefont {Bian}}, \bibinfo {author} {\bibfnamefont
  {C.}~\bibnamefont {Zhang}}, \bibinfo {author} {\bibfnamefont
  {R.}~\bibnamefont {Sankar}}, \bibinfo {author} {\bibfnamefont
  {G.}~\bibnamefont {Chang}}, \bibinfo {author} {\bibfnamefont
  {Z.}~\bibnamefont {Yuan}}, \bibinfo {author} {\bibfnamefont {C.-C.}\
  \bibnamefont {Lee}},  \emph {et~al.},\ }\href@noop {} {\bibfield  {journal}
  {\bibinfo  {journal} {Science}\ }\textbf {\bibinfo {volume} {349}},\ \bibinfo
  {pages} {613} (\bibinfo {year} {2015}{\natexlab{a}})}\BibitemShut {NoStop}%
\bibitem [{\citenamefont {Huang}\ \emph
  {et~al.}(2015{\natexlab{a}})\citenamefont {Huang}, \citenamefont {Xu},
  \citenamefont {Belopolski}, \citenamefont {Lee}, \citenamefont {Chang},
  \citenamefont {Wang}, \citenamefont {Alidoust}, \citenamefont {Bian},
  \citenamefont {Neupane}, \citenamefont {Zhang} \emph
  {et~al.}}]{huang2015weyl}%
  \BibitemOpen
  \bibfield  {author} {\bibinfo {author} {\bibfnamefont {S.-M.}\ \bibnamefont
  {Huang}}, \bibinfo {author} {\bibfnamefont {S.-Y.}\ \bibnamefont {Xu}},
  \bibinfo {author} {\bibfnamefont {I.}~\bibnamefont {Belopolski}}, \bibinfo
  {author} {\bibfnamefont {C.-C.}\ \bibnamefont {Lee}}, \bibinfo {author}
  {\bibfnamefont {G.}~\bibnamefont {Chang}}, \bibinfo {author} {\bibfnamefont
  {B.}~\bibnamefont {Wang}}, \bibinfo {author} {\bibfnamefont {N.}~\bibnamefont
  {Alidoust}}, \bibinfo {author} {\bibfnamefont {G.}~\bibnamefont {Bian}},
  \bibinfo {author} {\bibfnamefont {M.}~\bibnamefont {Neupane}}, \bibinfo
  {author} {\bibfnamefont {C.}~\bibnamefont {Zhang}},  \emph {et~al.},\
  }\href@noop {} {\bibfield  {journal} {\bibinfo  {journal} {Nat. Commun.}\
  }\textbf {\bibinfo {volume} {6}},\ \bibinfo {pages} {7373} (\bibinfo {year}
  {2015}{\natexlab{a}})}\BibitemShut {NoStop}%
\bibitem [{\citenamefont {Lv}\ \emph {et~al.}(2015{\natexlab{a}})\citenamefont
  {Lv}, \citenamefont {Weng}, \citenamefont {Fu}, \citenamefont {Wang},
  \citenamefont {Miao}, \citenamefont {Ma}, \citenamefont {Richard},
  \citenamefont {Huang}, \citenamefont {Zhao}, \citenamefont {Chen},
  \citenamefont {Fang}, \citenamefont {Dai}, \citenamefont {Qian},\ and\
  \citenamefont {Ding}}]{PhysRevX.5.031013DingHongTaAs}%
  \BibitemOpen
  \bibfield  {author} {\bibinfo {author} {\bibfnamefont {B.~Q.}\ \bibnamefont
  {Lv}}, \bibinfo {author} {\bibfnamefont {H.~M.}\ \bibnamefont {Weng}},
  \bibinfo {author} {\bibfnamefont {B.~B.}\ \bibnamefont {Fu}}, \bibinfo
  {author} {\bibfnamefont {X.~P.}\ \bibnamefont {Wang}}, \bibinfo {author}
  {\bibfnamefont {H.}~\bibnamefont {Miao}}, \bibinfo {author} {\bibfnamefont
  {J.}~\bibnamefont {Ma}}, \bibinfo {author} {\bibfnamefont {P.}~\bibnamefont
  {Richard}}, \bibinfo {author} {\bibfnamefont {X.~C.}\ \bibnamefont {Huang}},
  \bibinfo {author} {\bibfnamefont {L.~X.}\ \bibnamefont {Zhao}}, \bibinfo
  {author} {\bibfnamefont {G.~F.}\ \bibnamefont {Chen}}, \bibinfo {author}
  {\bibfnamefont {Z.}~\bibnamefont {Fang}}, \bibinfo {author} {\bibfnamefont
  {X.}~\bibnamefont {Dai}}, \bibinfo {author} {\bibfnamefont {T.}~\bibnamefont
  {Qian}}, \ and\ \bibinfo {author} {\bibfnamefont {H.}~\bibnamefont {Ding}},\
  }\href@noop {} {\bibfield  {journal} {\bibinfo  {journal} {Phys. Rev. X}\
  }\textbf {\bibinfo {volume} {5}},\ \bibinfo {pages} {031013} (\bibinfo {year}
  {2015}{\natexlab{a}})}\BibitemShut {NoStop}%
\bibitem [{\citenamefont {Lv}\ \emph {et~al.}(2015{\natexlab{b}})\citenamefont
  {Lv}, \citenamefont {Xu}, \citenamefont {Weng}, \citenamefont {Ma},
  \citenamefont {Richard}, \citenamefont {Huang}, \citenamefont {Zhao},
  \citenamefont {Chen}, \citenamefont {Matt}, \citenamefont {Bisti} \emph
  {et~al.}}]{NatPhysDingHongTaAs}%
  \BibitemOpen
  \bibfield  {author} {\bibinfo {author} {\bibfnamefont {B.}~\bibnamefont
  {Lv}}, \bibinfo {author} {\bibfnamefont {N.}~\bibnamefont {Xu}}, \bibinfo
  {author} {\bibfnamefont {H.}~\bibnamefont {Weng}}, \bibinfo {author}
  {\bibfnamefont {J.}~\bibnamefont {Ma}}, \bibinfo {author} {\bibfnamefont
  {P.}~\bibnamefont {Richard}}, \bibinfo {author} {\bibfnamefont
  {X.}~\bibnamefont {Huang}}, \bibinfo {author} {\bibfnamefont
  {L.}~\bibnamefont {Zhao}}, \bibinfo {author} {\bibfnamefont {G.}~\bibnamefont
  {Chen}}, \bibinfo {author} {\bibfnamefont {C.}~\bibnamefont {Matt}}, \bibinfo
  {author} {\bibfnamefont {F.}~\bibnamefont {Bisti}},  \emph {et~al.},\
  }\href@noop {} {\bibfield  {journal} {\bibinfo  {journal} {Nat. Phys.}\
  }\textbf {\bibinfo {volume} {11}},\ \bibinfo {pages} {724} (\bibinfo {year}
  {2015}{\natexlab{b}})}\BibitemShut {NoStop}%
\bibitem [{\citenamefont {Yang}\ \emph {et~al.}(2015)\citenamefont {Yang},
  \citenamefont {Liu}, \citenamefont {Sun}, \citenamefont {Peng}, \citenamefont
  {Yang}, \citenamefont {Zhang}, \citenamefont {Zhou}, \citenamefont {Zhang},
  \citenamefont {Guo}, \citenamefont {Rahn} \emph {et~al.}}]{yang2015weyl}%
  \BibitemOpen
  \bibfield  {author} {\bibinfo {author} {\bibfnamefont {L.}~\bibnamefont
  {Yang}}, \bibinfo {author} {\bibfnamefont {Z.}~\bibnamefont {Liu}}, \bibinfo
  {author} {\bibfnamefont {Y.}~\bibnamefont {Sun}}, \bibinfo {author}
  {\bibfnamefont {H.}~\bibnamefont {Peng}}, \bibinfo {author} {\bibfnamefont
  {H.}~\bibnamefont {Yang}}, \bibinfo {author} {\bibfnamefont {T.}~\bibnamefont
  {Zhang}}, \bibinfo {author} {\bibfnamefont {B.}~\bibnamefont {Zhou}},
  \bibinfo {author} {\bibfnamefont {Y.}~\bibnamefont {Zhang}}, \bibinfo
  {author} {\bibfnamefont {Y.}~\bibnamefont {Guo}}, \bibinfo {author}
  {\bibfnamefont {M.}~\bibnamefont {Rahn}},  \emph {et~al.},\ }\href@noop {}
  {\bibfield  {journal} {\bibinfo  {journal} {Nat. Phys.}\ }\textbf {\bibinfo
  {volume} {11}},\ \bibinfo {pages} {728} (\bibinfo {year} {2015})}\BibitemShut
  {NoStop}%
\bibitem [{\citenamefont {Xu}\ \emph {et~al.}(2015{\natexlab{b}})\citenamefont
  {Xu}, \citenamefont {Alidoust}, \citenamefont {Belopolski}, \citenamefont
  {Yuan}, \citenamefont {Bian}, \citenamefont {Chang}, \citenamefont {Zheng},
  \citenamefont {Strocov}, \citenamefont {Sanchez}, \citenamefont {Chang} \emph
  {et~al.}}]{xu2015discoveryNbAs}%
  \BibitemOpen
  \bibfield  {author} {\bibinfo {author} {\bibfnamefont {S.-Y.}\ \bibnamefont
  {Xu}}, \bibinfo {author} {\bibfnamefont {N.}~\bibnamefont {Alidoust}},
  \bibinfo {author} {\bibfnamefont {I.}~\bibnamefont {Belopolski}}, \bibinfo
  {author} {\bibfnamefont {Z.}~\bibnamefont {Yuan}}, \bibinfo {author}
  {\bibfnamefont {G.}~\bibnamefont {Bian}}, \bibinfo {author} {\bibfnamefont
  {T.-R.}\ \bibnamefont {Chang}}, \bibinfo {author} {\bibfnamefont
  {H.}~\bibnamefont {Zheng}}, \bibinfo {author} {\bibfnamefont {V.~N.}\
  \bibnamefont {Strocov}}, \bibinfo {author} {\bibfnamefont {D.~S.}\
  \bibnamefont {Sanchez}}, \bibinfo {author} {\bibfnamefont {G.}~\bibnamefont
  {Chang}},  \emph {et~al.},\ }\href@noop {} {\bibfield  {journal} {\bibinfo
  {journal} {Nat. Phys.}\ }\textbf {\bibinfo {volume} {11}},\ \bibinfo {pages}
  {748} (\bibinfo {year} {2015}{\natexlab{b}})}\BibitemShut {NoStop}%
\bibitem [{\citenamefont {Xu}\ \emph {et~al.}(2015{\natexlab{c}})\citenamefont
  {Xu}, \citenamefont {Belopolski}, \citenamefont {Sanchez}, \citenamefont
  {Zhang}, \citenamefont {Chang}, \citenamefont {Guo}, \citenamefont {Bian},
  \citenamefont {Yuan}, \citenamefont {Lu}, \citenamefont {Chang} \emph
  {et~al.}}]{xu2015experimental}%
  \BibitemOpen
  \bibfield  {author} {\bibinfo {author} {\bibfnamefont {S.-Y.}\ \bibnamefont
  {Xu}}, \bibinfo {author} {\bibfnamefont {I.}~\bibnamefont {Belopolski}},
  \bibinfo {author} {\bibfnamefont {D.~S.}\ \bibnamefont {Sanchez}}, \bibinfo
  {author} {\bibfnamefont {C.}~\bibnamefont {Zhang}}, \bibinfo {author}
  {\bibfnamefont {G.}~\bibnamefont {Chang}}, \bibinfo {author} {\bibfnamefont
  {C.}~\bibnamefont {Guo}}, \bibinfo {author} {\bibfnamefont {G.}~\bibnamefont
  {Bian}}, \bibinfo {author} {\bibfnamefont {Z.}~\bibnamefont {Yuan}}, \bibinfo
  {author} {\bibfnamefont {H.}~\bibnamefont {Lu}}, \bibinfo {author}
  {\bibfnamefont {T.-R.}\ \bibnamefont {Chang}},  \emph {et~al.},\ }\href@noop
  {} {\bibfield  {journal} {\bibinfo  {journal} {Sci. Adv.}\ }\textbf {\bibinfo
  {volume} {1}},\ \bibinfo {pages} {e1501092} (\bibinfo {year}
  {2015}{\natexlab{c}})}\BibitemShut {NoStop}%
\bibitem [{\citenamefont {Xu}\ \emph {et~al.}(2016)\citenamefont {Xu},
  \citenamefont {Weng}, \citenamefont {Lv}, \citenamefont {Matt}, \citenamefont
  {Park}, \citenamefont {Bisti}, \citenamefont {Strocov}, \citenamefont
  {Gawryluk}, \citenamefont {Pomjakushina}, \citenamefont {Conder} \emph
  {et~al.}}]{xu2016observation}%
  \BibitemOpen
  \bibfield  {author} {\bibinfo {author} {\bibfnamefont {N.}~\bibnamefont
  {Xu}}, \bibinfo {author} {\bibfnamefont {H.}~\bibnamefont {Weng}}, \bibinfo
  {author} {\bibfnamefont {B.}~\bibnamefont {Lv}}, \bibinfo {author}
  {\bibfnamefont {C.}~\bibnamefont {Matt}}, \bibinfo {author} {\bibfnamefont
  {J.}~\bibnamefont {Park}}, \bibinfo {author} {\bibfnamefont {F.}~\bibnamefont
  {Bisti}}, \bibinfo {author} {\bibfnamefont {V.}~\bibnamefont {Strocov}},
  \bibinfo {author} {\bibfnamefont {D.}~\bibnamefont {Gawryluk}}, \bibinfo
  {author} {\bibfnamefont {E.}~\bibnamefont {Pomjakushina}}, \bibinfo {author}
  {\bibfnamefont {K.}~\bibnamefont {Conder}},  \emph {et~al.},\ }\href@noop {}
  {\bibfield  {journal} {\bibinfo  {journal} {Nat. Commun.}\ }\textbf {\bibinfo
  {volume} {7}},\ \bibinfo {pages} {11006} (\bibinfo {year}
  {2016})}\BibitemShut {NoStop}%
\bibitem [{\citenamefont {Liu}\ \emph {et~al.}(2016{\natexlab{a}})\citenamefont
  {Liu}, \citenamefont {Yang}, \citenamefont {Sun}, \citenamefont {Zhang},
  \citenamefont {Peng}, \citenamefont {Yang}, \citenamefont {Chen},
  \citenamefont {Zhang}, \citenamefont {Guo}, \citenamefont {Prabhakaran} \emph
  {et~al.}}]{liu2016evolution}%
  \BibitemOpen
  \bibfield  {author} {\bibinfo {author} {\bibfnamefont {Z.}~\bibnamefont
  {Liu}}, \bibinfo {author} {\bibfnamefont {L.}~\bibnamefont {Yang}}, \bibinfo
  {author} {\bibfnamefont {Y.}~\bibnamefont {Sun}}, \bibinfo {author}
  {\bibfnamefont {T.}~\bibnamefont {Zhang}}, \bibinfo {author} {\bibfnamefont
  {H.}~\bibnamefont {Peng}}, \bibinfo {author} {\bibfnamefont {H.}~\bibnamefont
  {Yang}}, \bibinfo {author} {\bibfnamefont {C.}~\bibnamefont {Chen}}, \bibinfo
  {author} {\bibfnamefont {Y.}~\bibnamefont {Zhang}}, \bibinfo {author}
  {\bibfnamefont {Y.}~\bibnamefont {Guo}}, \bibinfo {author} {\bibfnamefont
  {D.}~\bibnamefont {Prabhakaran}},  \emph {et~al.},\ }\href@noop {} {\bibfield
   {journal} {\bibinfo  {journal} {Nat. Mater.}\ }\textbf {\bibinfo {volume}
  {15}},\ \bibinfo {pages} {27} (\bibinfo {year}
  {2016}{\natexlab{a}})}\BibitemShut {NoStop}%
\bibitem [{\citenamefont {Arnold}\ \emph {et~al.}(2016)\citenamefont {Arnold},
  \citenamefont {Shekhar}, \citenamefont {Wu}, \citenamefont {Sun},
  \citenamefont {Dos~Reis}, \citenamefont {Kumar}, \citenamefont {Naumann},
  \citenamefont {Ajeesh}, \citenamefont {Schmidt}, \citenamefont {Grushin}
  \emph {et~al.}}]{arnold2016negative}%
  \BibitemOpen
  \bibfield  {author} {\bibinfo {author} {\bibfnamefont {F.}~\bibnamefont
  {Arnold}}, \bibinfo {author} {\bibfnamefont {C.}~\bibnamefont {Shekhar}},
  \bibinfo {author} {\bibfnamefont {S.-C.}\ \bibnamefont {Wu}}, \bibinfo
  {author} {\bibfnamefont {Y.}~\bibnamefont {Sun}}, \bibinfo {author}
  {\bibfnamefont {R.~D.}\ \bibnamefont {Dos~Reis}}, \bibinfo {author}
  {\bibfnamefont {N.}~\bibnamefont {Kumar}}, \bibinfo {author} {\bibfnamefont
  {M.}~\bibnamefont {Naumann}}, \bibinfo {author} {\bibfnamefont {M.~O.}\
  \bibnamefont {Ajeesh}}, \bibinfo {author} {\bibfnamefont {M.}~\bibnamefont
  {Schmidt}}, \bibinfo {author} {\bibfnamefont {A.~G.}\ \bibnamefont
  {Grushin}},  \emph {et~al.},\ }\href@noop {} {\bibfield  {journal} {\bibinfo
  {journal} {Nat. Commun.}\ }\textbf {\bibinfo {volume} {7}},\ \bibinfo {pages}
  {11615} (\bibinfo {year} {2016})}\BibitemShut {NoStop}%
\bibitem [{\citenamefont {Zhang}\ \emph
  {et~al.}(2016{\natexlab{a}})\citenamefont {Zhang}, \citenamefont {Xu},
  \citenamefont {Belopolski}, \citenamefont {Yuan}, \citenamefont {Lin},
  \citenamefont {Tong}, \citenamefont {Bian}, \citenamefont {Alidoust},
  \citenamefont {Lee}, \citenamefont {Huang} \emph
  {et~al.}}]{zhang2016signatures}%
  \BibitemOpen
  \bibfield  {author} {\bibinfo {author} {\bibfnamefont {C.-L.}\ \bibnamefont
  {Zhang}}, \bibinfo {author} {\bibfnamefont {S.-Y.}\ \bibnamefont {Xu}},
  \bibinfo {author} {\bibfnamefont {I.}~\bibnamefont {Belopolski}}, \bibinfo
  {author} {\bibfnamefont {Z.}~\bibnamefont {Yuan}}, \bibinfo {author}
  {\bibfnamefont {Z.}~\bibnamefont {Lin}}, \bibinfo {author} {\bibfnamefont
  {B.}~\bibnamefont {Tong}}, \bibinfo {author} {\bibfnamefont {G.}~\bibnamefont
  {Bian}}, \bibinfo {author} {\bibfnamefont {N.}~\bibnamefont {Alidoust}},
  \bibinfo {author} {\bibfnamefont {C.-C.}\ \bibnamefont {Lee}}, \bibinfo
  {author} {\bibfnamefont {S.-M.}\ \bibnamefont {Huang}},  \emph {et~al.},\
  }\href@noop {} {\bibfield  {journal} {\bibinfo  {journal} {Nat. Commun.}\
  }\textbf {\bibinfo {volume} {7}},\ \bibinfo {pages} {10735} (\bibinfo {year}
  {2016}{\natexlab{a}})}\BibitemShut {NoStop}%
\bibitem [{\citenamefont {Huang}\ \emph
  {et~al.}(2015{\natexlab{b}})\citenamefont {Huang}, \citenamefont {Zhao},
  \citenamefont {Long}, \citenamefont {Wang}, \citenamefont {Chen},
  \citenamefont {Yang}, \citenamefont {Liang}, \citenamefont {Xue},
  \citenamefont {Weng}, \citenamefont {Fang}, \citenamefont {Dai},\ and\
  \citenamefont {Chen}}]{PhysRevX.5.031023}%
  \BibitemOpen
  \bibfield  {author} {\bibinfo {author} {\bibfnamefont {X.}~\bibnamefont
  {Huang}}, \bibinfo {author} {\bibfnamefont {L.}~\bibnamefont {Zhao}},
  \bibinfo {author} {\bibfnamefont {Y.}~\bibnamefont {Long}}, \bibinfo {author}
  {\bibfnamefont {P.}~\bibnamefont {Wang}}, \bibinfo {author} {\bibfnamefont
  {D.}~\bibnamefont {Chen}}, \bibinfo {author} {\bibfnamefont {Z.}~\bibnamefont
  {Yang}}, \bibinfo {author} {\bibfnamefont {H.}~\bibnamefont {Liang}},
  \bibinfo {author} {\bibfnamefont {M.}~\bibnamefont {Xue}}, \bibinfo {author}
  {\bibfnamefont {H.}~\bibnamefont {Weng}}, \bibinfo {author} {\bibfnamefont
  {Z.}~\bibnamefont {Fang}}, \bibinfo {author} {\bibfnamefont {X.}~\bibnamefont
  {Dai}}, \ and\ \bibinfo {author} {\bibfnamefont {G.}~\bibnamefont {Chen}},\
  }\href@noop {} {\bibfield  {journal} {\bibinfo  {journal} {Phys. Rev. X}\
  }\textbf {\bibinfo {volume} {5}},\ \bibinfo {pages} {031023} (\bibinfo {year}
  {2015}{\natexlab{b}})}\BibitemShut {NoStop}%
\bibitem [{\citenamefont {Hu}\ \emph {et~al.}(2015)\citenamefont {Hu},
  \citenamefont {Liu}, \citenamefont {Graf}, \citenamefont {Radmanesh},
  \citenamefont {Adams}, \citenamefont {Chuang}, \citenamefont {Wang},
  \citenamefont {Chiorescu}, \citenamefont {Wei}, \citenamefont {Spinu} \emph
  {et~al.}}]{hu2015pi}%
  \BibitemOpen
  \bibfield  {author} {\bibinfo {author} {\bibfnamefont {J.}~\bibnamefont
  {Hu}}, \bibinfo {author} {\bibfnamefont {J.}~\bibnamefont {Liu}}, \bibinfo
  {author} {\bibfnamefont {D.}~\bibnamefont {Graf}}, \bibinfo {author}
  {\bibfnamefont {S.}~\bibnamefont {Radmanesh}}, \bibinfo {author}
  {\bibfnamefont {D.}~\bibnamefont {Adams}}, \bibinfo {author} {\bibfnamefont
  {A.}~\bibnamefont {Chuang}}, \bibinfo {author} {\bibfnamefont
  {Y.}~\bibnamefont {Wang}}, \bibinfo {author} {\bibfnamefont {I.}~\bibnamefont
  {Chiorescu}}, \bibinfo {author} {\bibfnamefont {J.}~\bibnamefont {Wei}},
  \bibinfo {author} {\bibfnamefont {L.}~\bibnamefont {Spinu}},  \emph
  {et~al.},\ }\href@noop {} {\bibfield  {journal} {\bibinfo  {journal} {Sci
  Rep}\ }\textbf {\bibinfo {volume} {6}},\ \bibinfo {pages} {18674} (\bibinfo
  {year} {2015})}\BibitemShut {NoStop}%
\bibitem [{\citenamefont {Shekhar}\ \emph {et~al.}(2015)\citenamefont
  {Shekhar}, \citenamefont {Nayak}, \citenamefont {Sun}, \citenamefont
  {Schmidt}, \citenamefont {Nicklas}, \citenamefont {Leermakers}, \citenamefont
  {Zeitler}, \citenamefont {Skourski}, \citenamefont {Wosnitza},\ and\
  \citenamefont {Liu}}]{borrmann2015extremely}%
  \BibitemOpen
  \bibfield  {author} {\bibinfo {author} {\bibfnamefont {C.}~\bibnamefont
  {Shekhar}}, \bibinfo {author} {\bibfnamefont {A.~K.}\ \bibnamefont {Nayak}},
  \bibinfo {author} {\bibfnamefont {Y.}~\bibnamefont {Sun}}, \bibinfo {author}
  {\bibfnamefont {M.}~\bibnamefont {Schmidt}}, \bibinfo {author} {\bibfnamefont
  {M.}~\bibnamefont {Nicklas}}, \bibinfo {author} {\bibfnamefont
  {I.}~\bibnamefont {Leermakers}}, \bibinfo {author} {\bibfnamefont
  {U.}~\bibnamefont {Zeitler}}, \bibinfo {author} {\bibfnamefont
  {Y.}~\bibnamefont {Skourski}}, \bibinfo {author} {\bibfnamefont
  {J.}~\bibnamefont {Wosnitza}}, \ and\ \bibinfo {author} {\bibfnamefont
  {Z.}~\bibnamefont {Liu}},\ }\href@noop {} {\bibfield  {journal} {\bibinfo
  {journal} {Nat. Phys.}\ }\textbf {\bibinfo {volume} {11}},\ \bibinfo {pages}
  {645} (\bibinfo {year} {2015})}\BibitemShut {NoStop}%
\bibitem [{\citenamefont {Li}\ \emph {et~al.}(2015{\natexlab{a}})\citenamefont
  {Li}, \citenamefont {Wang}, \citenamefont {Liu}, \citenamefont {Wang},
  \citenamefont {Liao},\ and\ \citenamefont {Yu}}]{li2015giant}%
  \BibitemOpen
  \bibfield  {author} {\bibinfo {author} {\bibfnamefont {C.-Z.}\ \bibnamefont
  {Li}}, \bibinfo {author} {\bibfnamefont {L.-X.}\ \bibnamefont {Wang}},
  \bibinfo {author} {\bibfnamefont {H.}~\bibnamefont {Liu}}, \bibinfo {author}
  {\bibfnamefont {J.}~\bibnamefont {Wang}}, \bibinfo {author} {\bibfnamefont
  {Z.-M.}\ \bibnamefont {Liao}}, \ and\ \bibinfo {author} {\bibfnamefont
  {D.-P.}\ \bibnamefont {Yu}},\ }\href@noop {} {\bibfield  {journal} {\bibinfo
  {journal} {Nat. Commun.}\ }\textbf {\bibinfo {volume} {6}},\ \bibinfo {pages}
  {10137} (\bibinfo {year} {2015}{\natexlab{a}})}\BibitemShut {NoStop}%
\bibitem [{\citenamefont {Li}\ \emph {et~al.}(2015{\natexlab{b}})\citenamefont
  {Li}, \citenamefont {He}, \citenamefont {Lu}, \citenamefont {Zhang},
  \citenamefont {Liu}, \citenamefont {Ma}, \citenamefont {Fan}, \citenamefont
  {Shen},\ and\ \citenamefont {Wang}}]{li2015negative}%
  \BibitemOpen
  \bibfield  {author} {\bibinfo {author} {\bibfnamefont {H.}~\bibnamefont
  {Li}}, \bibinfo {author} {\bibfnamefont {H.}~\bibnamefont {He}}, \bibinfo
  {author} {\bibfnamefont {H.}~\bibnamefont {Lu}}, \bibinfo {author}
  {\bibfnamefont {H.}~\bibnamefont {Zhang}}, \bibinfo {author} {\bibfnamefont
  {H.}~\bibnamefont {Liu}}, \bibinfo {author} {\bibfnamefont {R.}~\bibnamefont
  {Ma}}, \bibinfo {author} {\bibfnamefont {Z.}~\bibnamefont {Fan}}, \bibinfo
  {author} {\bibfnamefont {S.}~\bibnamefont {Shen}}, \ and\ \bibinfo {author}
  {\bibfnamefont {J.}~\bibnamefont {Wang}},\ }\href@noop {} {\bibfield
  {journal} {\bibinfo  {journal} {Nat. Commun.}\ }\textbf {\bibinfo {volume}
  {7}},\ \bibinfo {pages} {10301} (\bibinfo {year}
  {2015}{\natexlab{b}})}\BibitemShut {NoStop}%
\bibitem [{\citenamefont {Xiong}\ \emph {et~al.}(2015)\citenamefont {Xiong},
  \citenamefont {Kushwaha}, \citenamefont {Liang}, \citenamefont {Krizan},
  \citenamefont {Hirschberger}, \citenamefont {Wang}, \citenamefont {Cava},\
  and\ \citenamefont {Ong}}]{xiong2015evidence}%
  \BibitemOpen
  \bibfield  {author} {\bibinfo {author} {\bibfnamefont {J.}~\bibnamefont
  {Xiong}}, \bibinfo {author} {\bibfnamefont {S.~K.}\ \bibnamefont {Kushwaha}},
  \bibinfo {author} {\bibfnamefont {T.}~\bibnamefont {Liang}}, \bibinfo
  {author} {\bibfnamefont {J.~W.}\ \bibnamefont {Krizan}}, \bibinfo {author}
  {\bibfnamefont {M.}~\bibnamefont {Hirschberger}}, \bibinfo {author}
  {\bibfnamefont {W.}~\bibnamefont {Wang}}, \bibinfo {author} {\bibfnamefont
  {R.}~\bibnamefont {Cava}}, \ and\ \bibinfo {author} {\bibfnamefont
  {N.}~\bibnamefont {Ong}},\ }\href@noop {} {\bibfield  {journal} {\bibinfo
  {journal} {Science}\ }\textbf {\bibinfo {volume} {350}},\ \bibinfo {pages}
  {413} (\bibinfo {year} {2015})}\BibitemShut {NoStop}%
\bibitem [{\citenamefont {Xiong}\ \emph {et~al.}(2016)\citenamefont {Xiong},
  \citenamefont {Kushwaha}, \citenamefont {Krizan}, \citenamefont {Liang},
  \citenamefont {Cava},\ and\ \citenamefont {Ong}}]{xiong2016anomalous}%
  \BibitemOpen
  \bibfield  {author} {\bibinfo {author} {\bibfnamefont {J.}~\bibnamefont
  {Xiong}}, \bibinfo {author} {\bibfnamefont {S.}~\bibnamefont {Kushwaha}},
  \bibinfo {author} {\bibfnamefont {J.}~\bibnamefont {Krizan}}, \bibinfo
  {author} {\bibfnamefont {T.}~\bibnamefont {Liang}}, \bibinfo {author}
  {\bibfnamefont {R.}~\bibnamefont {Cava}}, \ and\ \bibinfo {author}
  {\bibfnamefont {N.}~\bibnamefont {Ong}},\ }\href@noop {} {\bibfield
  {journal} {\bibinfo  {journal} {EPL}\ }\textbf {\bibinfo {volume} {114}},\
  \bibinfo {pages} {27002} (\bibinfo {year} {2016})}\BibitemShut {NoStop}%
\bibitem [{\citenamefont {Wang}\ \emph
  {et~al.}(2012{\natexlab{b}})\citenamefont {Wang}, \citenamefont {Graf},
  \citenamefont {Wang}, \citenamefont {Lei}, \citenamefont {Tozer},\ and\
  \citenamefont {Petrovic}}]{PhysRevB.85.041101}%
  \BibitemOpen
  \bibfield  {author} {\bibinfo {author} {\bibfnamefont {K.}~\bibnamefont
  {Wang}}, \bibinfo {author} {\bibfnamefont {D.}~\bibnamefont {Graf}}, \bibinfo
  {author} {\bibfnamefont {L.}~\bibnamefont {Wang}}, \bibinfo {author}
  {\bibfnamefont {H.}~\bibnamefont {Lei}}, \bibinfo {author} {\bibfnamefont
  {S.~W.}\ \bibnamefont {Tozer}}, \ and\ \bibinfo {author} {\bibfnamefont
  {C.}~\bibnamefont {Petrovic}},\ }\href {\doibase 10.1103/PhysRevB.85.041101}
  {\bibfield  {journal} {\bibinfo  {journal} {Phys. Rev. B}\ }\textbf {\bibinfo
  {volume} {85}},\ \bibinfo {pages} {041101} (\bibinfo {year}
  {2012}{\natexlab{b}})}\BibitemShut {NoStop}%
\bibitem [{\citenamefont {He}\ \emph {et~al.}(2012)\citenamefont {He},
  \citenamefont {Wang},\ and\ \citenamefont {Chen}}]{he2012giant}%
  \BibitemOpen
  \bibfield  {author} {\bibinfo {author} {\bibfnamefont {J.}~\bibnamefont
  {He}}, \bibinfo {author} {\bibfnamefont {D.}~\bibnamefont {Wang}}, \ and\
  \bibinfo {author} {\bibfnamefont {G.}~\bibnamefont {Chen}},\ }\href@noop {}
  {\bibfield  {journal} {\bibinfo  {journal} {Appl. Phys. Lett.}\ }\textbf
  {\bibinfo {volume} {100}},\ \bibinfo {pages} {112405} (\bibinfo {year}
  {2012})}\BibitemShut {NoStop}%
\bibitem [{\citenamefont {Feng}\ \emph {et~al.}(2014)\citenamefont {Feng},
  \citenamefont {Wang}, \citenamefont {Chen}, \citenamefont {Shi},
  \citenamefont {Xie}, \citenamefont {Yi}, \citenamefont {Liang}, \citenamefont
  {He}, \citenamefont {He}, \citenamefont {Peng} \emph
  {et~al.}}]{feng2014strong}%
  \BibitemOpen
  \bibfield  {author} {\bibinfo {author} {\bibfnamefont {Y.}~\bibnamefont
  {Feng}}, \bibinfo {author} {\bibfnamefont {Z.}~\bibnamefont {Wang}}, \bibinfo
  {author} {\bibfnamefont {C.}~\bibnamefont {Chen}}, \bibinfo {author}
  {\bibfnamefont {Y.}~\bibnamefont {Shi}}, \bibinfo {author} {\bibfnamefont
  {Z.}~\bibnamefont {Xie}}, \bibinfo {author} {\bibfnamefont {H.}~\bibnamefont
  {Yi}}, \bibinfo {author} {\bibfnamefont {A.}~\bibnamefont {Liang}}, \bibinfo
  {author} {\bibfnamefont {S.}~\bibnamefont {He}}, \bibinfo {author}
  {\bibfnamefont {J.}~\bibnamefont {He}}, \bibinfo {author} {\bibfnamefont
  {Y.}~\bibnamefont {Peng}},  \emph {et~al.},\ }\href@noop {} {\bibfield
  {journal} {\bibinfo  {journal} {Sci Rep}\ }\textbf {\bibinfo {volume} {4}},\
  \bibinfo {pages} {5385} (\bibinfo {year} {2014})}\BibitemShut {NoStop}%
\bibitem [{\citenamefont {Lee}\ \emph {et~al.}(2013)\citenamefont {Lee},
  \citenamefont {Farhan}, \citenamefont {Kim},\ and\ \citenamefont
  {Shim}}]{PhysRevB.87.245104}%
  \BibitemOpen
  \bibfield  {author} {\bibinfo {author} {\bibfnamefont {G.}~\bibnamefont
  {Lee}}, \bibinfo {author} {\bibfnamefont {M.~A.}\ \bibnamefont {Farhan}},
  \bibinfo {author} {\bibfnamefont {J.~S.}\ \bibnamefont {Kim}}, \ and\
  \bibinfo {author} {\bibfnamefont {J.~H.}\ \bibnamefont {Shim}},\ }\href
  {\doibase 10.1103/PhysRevB.87.245104} {\bibfield  {journal} {\bibinfo
  {journal} {Phys. Rev. B}\ }\textbf {\bibinfo {volume} {87}},\ \bibinfo
  {pages} {245104} (\bibinfo {year} {2013})}\BibitemShut {NoStop}%
\bibitem [{\citenamefont {Park}\ \emph {et~al.}(2011)\citenamefont {Park},
  \citenamefont {Lee}, \citenamefont {Wolff-Fabris}, \citenamefont {Koh},
  \citenamefont {Eom}, \citenamefont {Kim}, \citenamefont {Farhan},
  \citenamefont {Jo}, \citenamefont {Kim}, \citenamefont {Shim},\ and\
  \citenamefont {Kim}}]{PhysRevLett.107.126402}%
  \BibitemOpen
  \bibfield  {author} {\bibinfo {author} {\bibfnamefont {J.}~\bibnamefont
  {Park}}, \bibinfo {author} {\bibfnamefont {G.}~\bibnamefont {Lee}}, \bibinfo
  {author} {\bibfnamefont {F.}~\bibnamefont {Wolff-Fabris}}, \bibinfo {author}
  {\bibfnamefont {Y.~Y.}\ \bibnamefont {Koh}}, \bibinfo {author} {\bibfnamefont
  {M.~J.}\ \bibnamefont {Eom}}, \bibinfo {author} {\bibfnamefont {Y.~K.}\
  \bibnamefont {Kim}}, \bibinfo {author} {\bibfnamefont {M.~A.}\ \bibnamefont
  {Farhan}}, \bibinfo {author} {\bibfnamefont {Y.~J.}\ \bibnamefont {Jo}},
  \bibinfo {author} {\bibfnamefont {C.}~\bibnamefont {Kim}}, \bibinfo {author}
  {\bibfnamefont {J.~H.}\ \bibnamefont {Shim}}, \ and\ \bibinfo {author}
  {\bibfnamefont {J.~S.}\ \bibnamefont {Kim}},\ }\href {\doibase
  10.1103/PhysRevLett.107.126402} {\bibfield  {journal} {\bibinfo  {journal}
  {Phys. Rev. Lett.}\ }\textbf {\bibinfo {volume} {107}},\ \bibinfo {pages}
  {126402} (\bibinfo {year} {2011})}\BibitemShut {NoStop}%
\bibitem [{\citenamefont {Wang}\ \emph
  {et~al.}(2011{\natexlab{a}})\citenamefont {Wang}, \citenamefont {Graf},
  \citenamefont {Lei}, \citenamefont {Tozer},\ and\ \citenamefont
  {Petrovic}}]{PhysRevB.84.220401}%
  \BibitemOpen
  \bibfield  {author} {\bibinfo {author} {\bibfnamefont {K.}~\bibnamefont
  {Wang}}, \bibinfo {author} {\bibfnamefont {D.}~\bibnamefont {Graf}}, \bibinfo
  {author} {\bibfnamefont {H.}~\bibnamefont {Lei}}, \bibinfo {author}
  {\bibfnamefont {S.~W.}\ \bibnamefont {Tozer}}, \ and\ \bibinfo {author}
  {\bibfnamefont {C.}~\bibnamefont {Petrovic}},\ }\href {\doibase
  10.1103/PhysRevB.84.220401} {\bibfield  {journal} {\bibinfo  {journal} {Phys.
  Rev. B}\ }\textbf {\bibinfo {volume} {84}},\ \bibinfo {pages} {220401}
  (\bibinfo {year} {2011}{\natexlab{a}})}\BibitemShut {NoStop}%
\bibitem [{\citenamefont {Wang}\ \emph
  {et~al.}(2011{\natexlab{b}})\citenamefont {Wang}, \citenamefont {Zhao},
  \citenamefont {Yin}, \citenamefont {Kotliar}, \citenamefont {Kim},
  \citenamefont {Aronson},\ and\ \citenamefont {Morosan}}]{PhysRevB.84.064428}%
  \BibitemOpen
  \bibfield  {author} {\bibinfo {author} {\bibfnamefont {J.~K.}\ \bibnamefont
  {Wang}}, \bibinfo {author} {\bibfnamefont {L.~L.}\ \bibnamefont {Zhao}},
  \bibinfo {author} {\bibfnamefont {Q.}~\bibnamefont {Yin}}, \bibinfo {author}
  {\bibfnamefont {G.}~\bibnamefont {Kotliar}}, \bibinfo {author} {\bibfnamefont
  {M.~S.}\ \bibnamefont {Kim}}, \bibinfo {author} {\bibfnamefont {M.~C.}\
  \bibnamefont {Aronson}}, \ and\ \bibinfo {author} {\bibfnamefont
  {E.}~\bibnamefont {Morosan}},\ }\href {\doibase 10.1103/PhysRevB.84.064428}
  {\bibfield  {journal} {\bibinfo  {journal} {Phys. Rev. B}\ }\textbf {\bibinfo
  {volume} {84}},\ \bibinfo {pages} {064428} (\bibinfo {year}
  {2011}{\natexlab{b}})}\BibitemShut {NoStop}%
\bibitem [{\citenamefont {Guo}\ \emph {et~al.}(2014)\citenamefont {Guo},
  \citenamefont {Princep}, \citenamefont {Zhang}, \citenamefont {Manuel},
  \citenamefont {Khalyavin}, \citenamefont {Mazin}, \citenamefont {Shi},\ and\
  \citenamefont {Boothroyd}}]{PhysRevB.90.075120}%
  \BibitemOpen
  \bibfield  {author} {\bibinfo {author} {\bibfnamefont {Y.~F.}\ \bibnamefont
  {Guo}}, \bibinfo {author} {\bibfnamefont {A.~J.}\ \bibnamefont {Princep}},
  \bibinfo {author} {\bibfnamefont {X.}~\bibnamefont {Zhang}}, \bibinfo
  {author} {\bibfnamefont {P.}~\bibnamefont {Manuel}}, \bibinfo {author}
  {\bibfnamefont {D.}~\bibnamefont {Khalyavin}}, \bibinfo {author}
  {\bibfnamefont {I.~I.}\ \bibnamefont {Mazin}}, \bibinfo {author}
  {\bibfnamefont {Y.~G.}\ \bibnamefont {Shi}}, \ and\ \bibinfo {author}
  {\bibfnamefont {A.~T.}\ \bibnamefont {Boothroyd}},\ }\href {\doibase
  10.1103/PhysRevB.90.075120} {\bibfield  {journal} {\bibinfo  {journal} {Phys.
  Rev. B}\ }\textbf {\bibinfo {volume} {90}},\ \bibinfo {pages} {075120}
  (\bibinfo {year} {2014})}\BibitemShut {NoStop}%
\bibitem [{\citenamefont {Jo}\ \emph {et~al.}(2014)\citenamefont {Jo},
  \citenamefont {Park}, \citenamefont {Lee}, \citenamefont {Eom}, \citenamefont
  {Choi}, \citenamefont {Shim}, \citenamefont {Kang},\ and\ \citenamefont
  {Kim}}]{PhysRevLett.113.156602}%
  \BibitemOpen
  \bibfield  {author} {\bibinfo {author} {\bibfnamefont {Y.~J.}\ \bibnamefont
  {Jo}}, \bibinfo {author} {\bibfnamefont {J.}~\bibnamefont {Park}}, \bibinfo
  {author} {\bibfnamefont {G.}~\bibnamefont {Lee}}, \bibinfo {author}
  {\bibfnamefont {M.~J.}\ \bibnamefont {Eom}}, \bibinfo {author} {\bibfnamefont
  {E.~S.}\ \bibnamefont {Choi}}, \bibinfo {author} {\bibfnamefont {J.~H.}\
  \bibnamefont {Shim}}, \bibinfo {author} {\bibfnamefont {W.}~\bibnamefont
  {Kang}}, \ and\ \bibinfo {author} {\bibfnamefont {J.~S.}\ \bibnamefont
  {Kim}},\ }\href {\doibase 10.1103/PhysRevLett.113.156602} {\bibfield
  {journal} {\bibinfo  {journal} {Phys. Rev. Lett.}\ }\textbf {\bibinfo
  {volume} {113}},\ \bibinfo {pages} {156602} (\bibinfo {year}
  {2014})}\BibitemShut {NoStop}%
\bibitem [{\citenamefont {Jia}\ \emph {et~al.}(2014)\citenamefont {Jia},
  \citenamefont {Liu}, \citenamefont {Cai}, \citenamefont {Qian}, \citenamefont
  {Wang}, \citenamefont {Miao}, \citenamefont {Richard}, \citenamefont {Zhao},
  \citenamefont {Li}, \citenamefont {Wang}, \citenamefont {He}, \citenamefont
  {Shi}, \citenamefont {Chen}, \citenamefont {Ding},\ and\ \citenamefont
  {Wang}}]{PhysRevB.90.035133}%
  \BibitemOpen
  \bibfield  {author} {\bibinfo {author} {\bibfnamefont {L.-L.}\ \bibnamefont
  {Jia}}, \bibinfo {author} {\bibfnamefont {Z.-H.}\ \bibnamefont {Liu}},
  \bibinfo {author} {\bibfnamefont {Y.-P.}\ \bibnamefont {Cai}}, \bibinfo
  {author} {\bibfnamefont {T.}~\bibnamefont {Qian}}, \bibinfo {author}
  {\bibfnamefont {X.-P.}\ \bibnamefont {Wang}}, \bibinfo {author}
  {\bibfnamefont {H.}~\bibnamefont {Miao}}, \bibinfo {author} {\bibfnamefont
  {P.}~\bibnamefont {Richard}}, \bibinfo {author} {\bibfnamefont {Y.-G.}\
  \bibnamefont {Zhao}}, \bibinfo {author} {\bibfnamefont {Y.}~\bibnamefont
  {Li}}, \bibinfo {author} {\bibfnamefont {D.-M.}\ \bibnamefont {Wang}},
  \bibinfo {author} {\bibfnamefont {J.-B.}\ \bibnamefont {He}}, \bibinfo
  {author} {\bibfnamefont {M.}~\bibnamefont {Shi}}, \bibinfo {author}
  {\bibfnamefont {G.-F.}\ \bibnamefont {Chen}}, \bibinfo {author}
  {\bibfnamefont {H.}~\bibnamefont {Ding}}, \ and\ \bibinfo {author}
  {\bibfnamefont {S.-C.}\ \bibnamefont {Wang}},\ }\href {\doibase
  10.1103/PhysRevB.90.035133} {\bibfield  {journal} {\bibinfo  {journal} {Phys.
  Rev. B}\ }\textbf {\bibinfo {volume} {90}},\ \bibinfo {pages} {035133}
  (\bibinfo {year} {2014})}\BibitemShut {NoStop}%
\bibitem [{\citenamefont {Zhang}\ \emph
  {et~al.}(2016{\natexlab{b}})\citenamefont {Zhang}, \citenamefont {Liu},
  \citenamefont {Yi}, \citenamefont {Zhao}, \citenamefont {Xia}, \citenamefont
  {Ji}, \citenamefont {Shi}, \citenamefont {Yu}, \citenamefont {Wang},
  \citenamefont {Chen} \emph {et~al.}}]{zhang2016interplay}%
  \BibitemOpen
  \bibfield  {author} {\bibinfo {author} {\bibfnamefont {A.}~\bibnamefont
  {Zhang}}, \bibinfo {author} {\bibfnamefont {C.}~\bibnamefont {Liu}}, \bibinfo
  {author} {\bibfnamefont {C.}~\bibnamefont {Yi}}, \bibinfo {author}
  {\bibfnamefont {G.}~\bibnamefont {Zhao}}, \bibinfo {author} {\bibfnamefont
  {T.-l.}\ \bibnamefont {Xia}}, \bibinfo {author} {\bibfnamefont
  {J.}~\bibnamefont {Ji}}, \bibinfo {author} {\bibfnamefont {Y.}~\bibnamefont
  {Shi}}, \bibinfo {author} {\bibfnamefont {R.}~\bibnamefont {Yu}}, \bibinfo
  {author} {\bibfnamefont {X.}~\bibnamefont {Wang}}, \bibinfo {author}
  {\bibfnamefont {C.}~\bibnamefont {Chen}},  \emph {et~al.},\ }\href@noop {}
  {\bibfield  {journal} {\bibinfo  {journal} {Nat. Commun.}\ }\textbf {\bibinfo
  {volume} {7}},\ \bibinfo {pages} {13833} (\bibinfo {year}
  {2016}{\natexlab{b}})}\BibitemShut {NoStop}%
\bibitem [{\citenamefont {Li}\ \emph {et~al.}(2016)\citenamefont {Li},
  \citenamefont {Wang}, \citenamefont {Graf}, \citenamefont {Wang},
  \citenamefont {Wang},\ and\ \citenamefont {Petrovic}}]{Petrovic2016}%
  \BibitemOpen
  \bibfield  {author} {\bibinfo {author} {\bibfnamefont {L.}~\bibnamefont
  {Li}}, \bibinfo {author} {\bibfnamefont {K.}~\bibnamefont {Wang}}, \bibinfo
  {author} {\bibfnamefont {D.}~\bibnamefont {Graf}}, \bibinfo {author}
  {\bibfnamefont {L.}~\bibnamefont {Wang}}, \bibinfo {author} {\bibfnamefont
  {A.}~\bibnamefont {Wang}}, \ and\ \bibinfo {author} {\bibfnamefont
  {C.}~\bibnamefont {Petrovic}},\ }\href {\doibase 10.1103/PhysRevB.93.115141}
  {\bibfield  {journal} {\bibinfo  {journal} {Phys. Rev. B}\ }\textbf {\bibinfo
  {volume} {93}},\ \bibinfo {pages} {115141} (\bibinfo {year}
  {2016})}\BibitemShut {NoStop}%
\bibitem [{\citenamefont {Wang}\ \emph
  {et~al.}(2016{\natexlab{a}})\citenamefont {Wang}, \citenamefont {Yu},\ and\
  \citenamefont {Xia}}]{wang2016large}%
  \BibitemOpen
  \bibfield  {author} {\bibinfo {author} {\bibfnamefont {Y.-Y.}\ \bibnamefont
  {Wang}}, \bibinfo {author} {\bibfnamefont {Q.-H.}\ \bibnamefont {Yu}}, \ and\
  \bibinfo {author} {\bibfnamefont {T.-L.}\ \bibnamefont {Xia}},\ }\href@noop
  {} {\bibfield  {journal} {\bibinfo  {journal} {Chin. Phys. B}\ }\textbf
  {\bibinfo {volume} {25}},\ \bibinfo {pages} {107503} (\bibinfo {year}
  {2016}{\natexlab{a}})}\BibitemShut {NoStop}%
\bibitem [{\citenamefont {Masuda}\ \emph {et~al.}(2016)\citenamefont {Masuda},
  \citenamefont {Sakai}, \citenamefont {Tokunaga}, \citenamefont {Yamasaki},
  \citenamefont {Miyake}, \citenamefont {Shiogai}, \citenamefont {Nakamura},
  \citenamefont {Awaji}, \citenamefont {Tsukazaki}, \citenamefont {Nakao} \emph
  {et~al.}}]{masuda2016quantum}%
  \BibitemOpen
  \bibfield  {author} {\bibinfo {author} {\bibfnamefont {H.}~\bibnamefont
  {Masuda}}, \bibinfo {author} {\bibfnamefont {H.}~\bibnamefont {Sakai}},
  \bibinfo {author} {\bibfnamefont {M.}~\bibnamefont {Tokunaga}}, \bibinfo
  {author} {\bibfnamefont {Y.}~\bibnamefont {Yamasaki}}, \bibinfo {author}
  {\bibfnamefont {A.}~\bibnamefont {Miyake}}, \bibinfo {author} {\bibfnamefont
  {J.}~\bibnamefont {Shiogai}}, \bibinfo {author} {\bibfnamefont
  {S.}~\bibnamefont {Nakamura}}, \bibinfo {author} {\bibfnamefont
  {S.}~\bibnamefont {Awaji}}, \bibinfo {author} {\bibfnamefont
  {A.}~\bibnamefont {Tsukazaki}}, \bibinfo {author} {\bibfnamefont
  {H.}~\bibnamefont {Nakao}},  \emph {et~al.},\ }\href@noop {} {\bibfield
  {journal} {\bibinfo  {journal} {Sci. Adv.}\ }\textbf {\bibinfo {volume}
  {2}},\ \bibinfo {pages} {e1501117} (\bibinfo {year} {2016})}\BibitemShut
  {NoStop}%
\bibitem [{\citenamefont {May}\ \emph {et~al.}(2014)\citenamefont {May},
  \citenamefont {McGuire},\ and\ \citenamefont {Sales}}]{PhysRevB.90.075109}%
  \BibitemOpen
  \bibfield  {author} {\bibinfo {author} {\bibfnamefont {A.~F.}\ \bibnamefont
  {May}}, \bibinfo {author} {\bibfnamefont {M.~A.}\ \bibnamefont {McGuire}}, \
  and\ \bibinfo {author} {\bibfnamefont {B.~C.}\ \bibnamefont {Sales}},\ }\href
  {\doibase 10.1103/PhysRevB.90.075109} {\bibfield  {journal} {\bibinfo
  {journal} {Phys. Rev. B}\ }\textbf {\bibinfo {volume} {90}},\ \bibinfo
  {pages} {075109} (\bibinfo {year} {2014})}\BibitemShut {NoStop}%
\bibitem [{\citenamefont {Borisenko}\ \emph {et~al.}(2015)\citenamefont
  {Borisenko}, \citenamefont {Evtushinsky}, \citenamefont {Gibson},
  \citenamefont {Yaresko}, \citenamefont {Kim}, \citenamefont {Ali},
  \citenamefont {Buechner}, \citenamefont {Hoesch},\ and\ \citenamefont
  {Cava}}]{borisenko2015time}%
  \BibitemOpen
  \bibfield  {author} {\bibinfo {author} {\bibfnamefont {S.}~\bibnamefont
  {Borisenko}}, \bibinfo {author} {\bibfnamefont {D.}~\bibnamefont
  {Evtushinsky}}, \bibinfo {author} {\bibfnamefont {Q.}~\bibnamefont {Gibson}},
  \bibinfo {author} {\bibfnamefont {A.}~\bibnamefont {Yaresko}}, \bibinfo
  {author} {\bibfnamefont {T.}~\bibnamefont {Kim}}, \bibinfo {author}
  {\bibfnamefont {M.}~\bibnamefont {Ali}}, \bibinfo {author} {\bibfnamefont
  {B.}~\bibnamefont {Buechner}}, \bibinfo {author} {\bibfnamefont
  {M.}~\bibnamefont {Hoesch}}, \ and\ \bibinfo {author} {\bibfnamefont {R.~J.}\
  \bibnamefont {Cava}},\ }\href@noop {} {\bibfield  {journal} {\bibinfo
  {journal} {arXiv preprint arXiv:1507.04847}\ } (\bibinfo {year}
  {2015})}\BibitemShut {NoStop}%
\bibitem [{\citenamefont {Wang}\ \emph
  {et~al.}(2016{\natexlab{b}})\citenamefont {Wang}, \citenamefont {Zaliznyak},
  \citenamefont {Ren}, \citenamefont {Wu}, \citenamefont {Graf}, \citenamefont
  {Garlea}, \citenamefont {Warren}, \citenamefont {Bozin}, \citenamefont
  {Zhu},\ and\ \citenamefont {Petrovic}}]{PhysRevB.94.165161}%
  \BibitemOpen
  \bibfield  {author} {\bibinfo {author} {\bibfnamefont {A.}~\bibnamefont
  {Wang}}, \bibinfo {author} {\bibfnamefont {I.}~\bibnamefont {Zaliznyak}},
  \bibinfo {author} {\bibfnamefont {W.}~\bibnamefont {Ren}}, \bibinfo {author}
  {\bibfnamefont {L.}~\bibnamefont {Wu}}, \bibinfo {author} {\bibfnamefont
  {D.}~\bibnamefont {Graf}}, \bibinfo {author} {\bibfnamefont {V.~O.}\
  \bibnamefont {Garlea}}, \bibinfo {author} {\bibfnamefont {J.~B.}\
  \bibnamefont {Warren}}, \bibinfo {author} {\bibfnamefont {E.}~\bibnamefont
  {Bozin}}, \bibinfo {author} {\bibfnamefont {Y.}~\bibnamefont {Zhu}}, \ and\
  \bibinfo {author} {\bibfnamefont {C.}~\bibnamefont {Petrovic}},\ }\href
  {\doibase 10.1103/PhysRevB.94.165161} {\bibfield  {journal} {\bibinfo
  {journal} {Phys. Rev. B}\ }\textbf {\bibinfo {volume} {94}},\ \bibinfo
  {pages} {165161} (\bibinfo {year} {2016}{\natexlab{b}})}\BibitemShut
  {NoStop}%
\bibitem [{\citenamefont {Farhan}\ \emph {et~al.}(2014)\citenamefont {Farhan},
  \citenamefont {Lee},\ and\ \citenamefont {Shim}}]{0953-8984-26-4-042201}%
  \BibitemOpen
  \bibfield  {author} {\bibinfo {author} {\bibfnamefont {M.~A.}\ \bibnamefont
  {Farhan}}, \bibinfo {author} {\bibfnamefont {G.}~\bibnamefont {Lee}}, \ and\
  \bibinfo {author} {\bibfnamefont {J.~H.}\ \bibnamefont {Shim}},\ }\href@noop
  {} {\bibfield  {journal} {\bibinfo  {journal} {J. Phys.-Condes. Matter}\
  }\textbf {\bibinfo {volume} {26}},\ \bibinfo {pages} {042201} (\bibinfo
  {year} {2014})}\BibitemShut {NoStop}%
\bibitem [{\citenamefont {Liu}\ \emph {et~al.}(2017)\citenamefont {Liu},
  \citenamefont {Hu}, \citenamefont {Zhang}, \citenamefont {Graf},
  \citenamefont {Cao}, \citenamefont {Radmanesh}, \citenamefont {Adams},
  \citenamefont {Zhu}, \citenamefont {Cheng}, \citenamefont {Liu} \emph
  {et~al.}}]{liu2017discovery}%
  \BibitemOpen
  \bibfield  {author} {\bibinfo {author} {\bibfnamefont {J.}~\bibnamefont
  {Liu}}, \bibinfo {author} {\bibfnamefont {J.}~\bibnamefont {Hu}}, \bibinfo
  {author} {\bibfnamefont {Q.}~\bibnamefont {Zhang}}, \bibinfo {author}
  {\bibfnamefont {D.}~\bibnamefont {Graf}}, \bibinfo {author} {\bibfnamefont
  {H.}~\bibnamefont {Cao}}, \bibinfo {author} {\bibfnamefont {S.}~\bibnamefont
  {Radmanesh}}, \bibinfo {author} {\bibfnamefont {D.}~\bibnamefont {Adams}},
  \bibinfo {author} {\bibfnamefont {Y.}~\bibnamefont {Zhu}}, \bibinfo {author}
  {\bibfnamefont {G.}~\bibnamefont {Cheng}}, \bibinfo {author} {\bibfnamefont
  {X.}~\bibnamefont {Liu}},  \emph {et~al.},\ }\href {\doibase
  10.1038/nmat4953} {\bibfield  {journal} {\bibinfo  {journal} {Nat. Mater.}\ }
  (\bibinfo {year} {2017}),\ 10.1038/nmat4953}\BibitemShut {NoStop}%
\bibitem [{\citenamefont {Liu}\ \emph {et~al.}(2016{\natexlab{b}})\citenamefont
  {Liu}, \citenamefont {Hu}, \citenamefont {Cao}, \citenamefont {Zhu},
  \citenamefont {Chuang}, \citenamefont {Graf}, \citenamefont {Adams},
  \citenamefont {Radmanesh}, \citenamefont {Spinu}, \citenamefont {Chiorescu}
  \emph {et~al.}}]{liu2016nearly}%
  \BibitemOpen
  \bibfield  {author} {\bibinfo {author} {\bibfnamefont {J.}~\bibnamefont
  {Liu}}, \bibinfo {author} {\bibfnamefont {J.}~\bibnamefont {Hu}}, \bibinfo
  {author} {\bibfnamefont {H.}~\bibnamefont {Cao}}, \bibinfo {author}
  {\bibfnamefont {Y.}~\bibnamefont {Zhu}}, \bibinfo {author} {\bibfnamefont
  {A.}~\bibnamefont {Chuang}}, \bibinfo {author} {\bibfnamefont
  {D.}~\bibnamefont {Graf}}, \bibinfo {author} {\bibfnamefont {D.}~\bibnamefont
  {Adams}}, \bibinfo {author} {\bibfnamefont {S.}~\bibnamefont {Radmanesh}},
  \bibinfo {author} {\bibfnamefont {L.}~\bibnamefont {Spinu}}, \bibinfo
  {author} {\bibfnamefont {I.}~\bibnamefont {Chiorescu}},  \emph {et~al.},\
  }\href@noop {} {\bibfield  {journal} {\bibinfo  {journal} {Sci Rep}\ }\textbf
  {\bibinfo {volume} {6}},\ \bibinfo {pages} {30525} (\bibinfo {year}
  {2016}{\natexlab{b}})}\BibitemShut {NoStop}%
\bibitem [{\citenamefont {He}\ \emph {et~al.}(2017)\citenamefont {He},
  \citenamefont {Fu}, \citenamefont {Zhao}, \citenamefont {Liang},
  \citenamefont {Chen}, \citenamefont {Leng}, \citenamefont {Wang},
  \citenamefont {Li}, \citenamefont {Zhang}, \citenamefont {Xue}, \citenamefont
  {Li}, \citenamefont {Zhang}, \citenamefont {Ren},\ and\ \citenamefont
  {Chen}}]{PhysRevB.95.045128}%
  \BibitemOpen
  \bibfield  {author} {\bibinfo {author} {\bibfnamefont {J.~B.}\ \bibnamefont
  {He}}, \bibinfo {author} {\bibfnamefont {Y.}~\bibnamefont {Fu}}, \bibinfo
  {author} {\bibfnamefont {L.~X.}\ \bibnamefont {Zhao}}, \bibinfo {author}
  {\bibfnamefont {H.}~\bibnamefont {Liang}}, \bibinfo {author} {\bibfnamefont
  {D.}~\bibnamefont {Chen}}, \bibinfo {author} {\bibfnamefont {Y.~M.}\
  \bibnamefont {Leng}}, \bibinfo {author} {\bibfnamefont {X.~M.}\ \bibnamefont
  {Wang}}, \bibinfo {author} {\bibfnamefont {J.}~\bibnamefont {Li}}, \bibinfo
  {author} {\bibfnamefont {S.}~\bibnamefont {Zhang}}, \bibinfo {author}
  {\bibfnamefont {M.~Q.}\ \bibnamefont {Xue}}, \bibinfo {author} {\bibfnamefont
  {C.~H.}\ \bibnamefont {Li}}, \bibinfo {author} {\bibfnamefont
  {P.}~\bibnamefont {Zhang}}, \bibinfo {author} {\bibfnamefont {Z.~A.}\
  \bibnamefont {Ren}}, \ and\ \bibinfo {author} {\bibfnamefont {G.~F.}\
  \bibnamefont {Chen}},\ }\href {\doibase 10.1103/PhysRevB.95.045128}
  {\bibfield  {journal} {\bibinfo  {journal} {Phys. Rev. B}\ }\textbf {\bibinfo
  {volume} {95}},\ \bibinfo {pages} {045128} (\bibinfo {year}
  {2017})}\BibitemShut {NoStop}%
\bibitem [{\citenamefont {Kealhofer}\ \emph {et~al.}(2017)\citenamefont
  {Kealhofer}, \citenamefont {Jang}, \citenamefont {Griffin}, \citenamefont
  {John}, \citenamefont {Benavides}, \citenamefont {Doyle}, \citenamefont
  {Helm}, \citenamefont {Moll}, \citenamefont {Neaton}, \citenamefont {Chan}
  \emph {et~al.}}]{kealhofer2017observation}%
  \BibitemOpen
  \bibfield  {author} {\bibinfo {author} {\bibfnamefont {R.}~\bibnamefont
  {Kealhofer}}, \bibinfo {author} {\bibfnamefont {S.}~\bibnamefont {Jang}},
  \bibinfo {author} {\bibfnamefont {S.~M.}\ \bibnamefont {Griffin}}, \bibinfo
  {author} {\bibfnamefont {C.}~\bibnamefont {John}}, \bibinfo {author}
  {\bibfnamefont {K.~A.}\ \bibnamefont {Benavides}}, \bibinfo {author}
  {\bibfnamefont {S.}~\bibnamefont {Doyle}}, \bibinfo {author} {\bibfnamefont
  {T.}~\bibnamefont {Helm}}, \bibinfo {author} {\bibfnamefont {P.~J.}\
  \bibnamefont {Moll}}, \bibinfo {author} {\bibfnamefont {J.~B.}\ \bibnamefont
  {Neaton}}, \bibinfo {author} {\bibfnamefont {J.~Y.}\ \bibnamefont {Chan}},
  \emph {et~al.},\ }\href@noop {} {\bibfield  {journal} {\bibinfo  {journal}
  {arXiv preprint arXiv:1708.03308}\ } (\bibinfo {year} {2017})}\BibitemShut
  {NoStop}%
\bibitem [{\citenamefont {Wang}\ \emph {et~al.}(2017)\citenamefont {Wang},
  \citenamefont {Xu}, \citenamefont {Sun},\ and\ \citenamefont
  {Xia}}]{wang2017Quantum}%
  \BibitemOpen
  \bibfield  {author} {\bibinfo {author} {\bibfnamefont {Y.-Y.}\ \bibnamefont
  {Wang}}, \bibinfo {author} {\bibfnamefont {S.}~\bibnamefont {Xu}}, \bibinfo
  {author} {\bibfnamefont {L.-L.}\ \bibnamefont {Sun}}, \ and\ \bibinfo
  {author} {\bibfnamefont {T.-L.}\ \bibnamefont {Xia}},\ }\href@noop {}
  {\bibfield  {journal} {\bibinfo  {journal} {arXiv preprint arXiv:1708.03913}\
  } (\bibinfo {year} {2017})}\BibitemShut {NoStop}%
\bibitem [{\citenamefont {Yang}\ \emph {et~al.}(1999)\citenamefont {Yang},
  \citenamefont {Liu}, \citenamefont {Hong}, \citenamefont {Reich},
  \citenamefont {Searson},\ and\ \citenamefont {Chien}}]{yang1999large}%
  \BibitemOpen
  \bibfield  {author} {\bibinfo {author} {\bibfnamefont {F.}~\bibnamefont
  {Yang}}, \bibinfo {author} {\bibfnamefont {K.}~\bibnamefont {Liu}}, \bibinfo
  {author} {\bibfnamefont {K.}~\bibnamefont {Hong}}, \bibinfo {author}
  {\bibfnamefont {D.}~\bibnamefont {Reich}}, \bibinfo {author} {\bibfnamefont
  {P.}~\bibnamefont {Searson}}, \ and\ \bibinfo {author} {\bibfnamefont
  {C.}~\bibnamefont {Chien}},\ }\href@noop {} {\bibfield  {journal} {\bibinfo
  {journal} {Science}\ }\textbf {\bibinfo {volume} {284}},\ \bibinfo {pages}
  {1335} (\bibinfo {year} {1999})}\BibitemShut {NoStop}%
\bibitem [{\citenamefont {Xu}\ \emph {et~al.}(1997)\citenamefont {Xu},
  \citenamefont {Husmann}, \citenamefont {Rosenbaum}, \citenamefont {Saboungi},
  \citenamefont {Enderby},\ and\ \citenamefont {Littlewood}}]{xu1997large}%
  \BibitemOpen
  \bibfield  {author} {\bibinfo {author} {\bibfnamefont {R.}~\bibnamefont
  {Xu}}, \bibinfo {author} {\bibfnamefont {A.}~\bibnamefont {Husmann}},
  \bibinfo {author} {\bibfnamefont {T.}~\bibnamefont {Rosenbaum}}, \bibinfo
  {author} {\bibfnamefont {M.-L.}\ \bibnamefont {Saboungi}}, \bibinfo {author}
  {\bibfnamefont {J.}~\bibnamefont {Enderby}}, \ and\ \bibinfo {author}
  {\bibfnamefont {P.}~\bibnamefont {Littlewood}},\ }\href@noop {} {\bibfield
  {journal} {\bibinfo  {journal} {Nature}\ }\textbf {\bibinfo {volume} {390}},\
  \bibinfo {pages} {57} (\bibinfo {year} {1997})}\BibitemShut {NoStop}%
\bibitem [{\citenamefont {Lee}\ \emph {et~al.}(2002)\citenamefont {Lee},
  \citenamefont {Rosenbaum}, \citenamefont {Saboungi},\ and\ \citenamefont
  {Schnyders}}]{PhysRevLett.88.066602}%
  \BibitemOpen
  \bibfield  {author} {\bibinfo {author} {\bibfnamefont {M.}~\bibnamefont
  {Lee}}, \bibinfo {author} {\bibfnamefont {T.~F.}\ \bibnamefont {Rosenbaum}},
  \bibinfo {author} {\bibfnamefont {M.-L.}\ \bibnamefont {Saboungi}}, \ and\
  \bibinfo {author} {\bibfnamefont {H.~S.}\ \bibnamefont {Schnyders}},\ }\href
  {\doibase 10.1103/PhysRevLett.88.066602} {\bibfield  {journal} {\bibinfo
  {journal} {Phys. Rev. Lett.}\ }\textbf {\bibinfo {volume} {88}},\ \bibinfo
  {pages} {066602} (\bibinfo {year} {2002})}\BibitemShut {NoStop}%
\bibitem [{\citenamefont {Hu}\ \emph {et~al.}(2005)\citenamefont {Hu},
  \citenamefont {Rosenbaum},\ and\ \citenamefont
  {Betts}}]{PhysRevLett.95.186603}%
  \BibitemOpen
  \bibfield  {author} {\bibinfo {author} {\bibfnamefont {J.}~\bibnamefont
  {Hu}}, \bibinfo {author} {\bibfnamefont {T.~F.}\ \bibnamefont {Rosenbaum}}, \
  and\ \bibinfo {author} {\bibfnamefont {J.~B.}\ \bibnamefont {Betts}},\ }\href
  {\doibase 10.1103/PhysRevLett.95.186603} {\bibfield  {journal} {\bibinfo
  {journal} {Phys. Rev. Lett.}\ }\textbf {\bibinfo {volume} {95}},\ \bibinfo
  {pages} {186603} (\bibinfo {year} {2005})}\BibitemShut {NoStop}%
\bibitem [{\citenamefont {von Kreutzbruck}\ \emph {et~al.}(2009)\citenamefont
  {von Kreutzbruck}, \citenamefont {Lembke}, \citenamefont {Mogwitz},
  \citenamefont {Korte},\ and\ \citenamefont {Janek}}]{PhysRevB.79.035204}%
  \BibitemOpen
  \bibfield  {author} {\bibinfo {author} {\bibfnamefont {M.}~\bibnamefont {von
  Kreutzbruck}}, \bibinfo {author} {\bibfnamefont {G.}~\bibnamefont {Lembke}},
  \bibinfo {author} {\bibfnamefont {B.}~\bibnamefont {Mogwitz}}, \bibinfo
  {author} {\bibfnamefont {C.}~\bibnamefont {Korte}}, \ and\ \bibinfo {author}
  {\bibfnamefont {J.}~\bibnamefont {Janek}},\ }\href {\doibase
  10.1103/PhysRevB.79.035204} {\bibfield  {journal} {\bibinfo  {journal} {Phys.
  Rev. B}\ }\textbf {\bibinfo {volume} {79}},\ \bibinfo {pages} {035204}
  (\bibinfo {year} {2009})}\BibitemShut {NoStop}%
\bibitem [{\citenamefont {Zhang}\ \emph {et~al.}(2011)\citenamefont {Zhang},
  \citenamefont {Yu}, \citenamefont {Feng}, \citenamefont {Yao}, \citenamefont
  {Weng}, \citenamefont {Dai},\ and\ \citenamefont
  {Fang}}]{PhysRevLett.106.156808}%
  \BibitemOpen
  \bibfield  {author} {\bibinfo {author} {\bibfnamefont {W.}~\bibnamefont
  {Zhang}}, \bibinfo {author} {\bibfnamefont {R.}~\bibnamefont {Yu}}, \bibinfo
  {author} {\bibfnamefont {W.}~\bibnamefont {Feng}}, \bibinfo {author}
  {\bibfnamefont {Y.}~\bibnamefont {Yao}}, \bibinfo {author} {\bibfnamefont
  {H.}~\bibnamefont {Weng}}, \bibinfo {author} {\bibfnamefont {X.}~\bibnamefont
  {Dai}}, \ and\ \bibinfo {author} {\bibfnamefont {Z.}~\bibnamefont {Fang}},\
  }\href {\doibase 10.1103/PhysRevLett.106.156808} {\bibfield  {journal}
  {\bibinfo  {journal} {Phys. Rev. Lett.}\ }\textbf {\bibinfo {volume} {106}},\
  \bibinfo {pages} {156808} (\bibinfo {year} {2011})}\BibitemShut {NoStop}%
\bibitem [{\citenamefont {Hu}\ and\ \citenamefont
  {Rosenbaum}(2008)}]{hu2008classical}%
  \BibitemOpen
  \bibfield  {author} {\bibinfo {author} {\bibfnamefont {J.}~\bibnamefont
  {Hu}}\ and\ \bibinfo {author} {\bibfnamefont {T.}~\bibnamefont {Rosenbaum}},\
  }\href@noop {} {\bibfield  {journal} {\bibinfo  {journal} {Nat. Mater.}\
  }\textbf {\bibinfo {volume} {7}},\ \bibinfo {pages} {697} (\bibinfo {year}
  {2008})}\BibitemShut {NoStop}%
\bibitem [{\citenamefont {Friedman}\ \emph {et~al.}(2010)\citenamefont
  {Friedman}, \citenamefont {Tedesco}, \citenamefont {Campbell}, \citenamefont
  {Culbertson}, \citenamefont {Aifer}, \citenamefont {Perkins}, \citenamefont
  {Myers-Ward}, \citenamefont {Hite}, \citenamefont {Eddy~Jr}, \citenamefont
  {Jernigan} \emph {et~al.}}]{friedman2010quantum}%
  \BibitemOpen
  \bibfield  {author} {\bibinfo {author} {\bibfnamefont {A.~L.}\ \bibnamefont
  {Friedman}}, \bibinfo {author} {\bibfnamefont {J.~L.}\ \bibnamefont
  {Tedesco}}, \bibinfo {author} {\bibfnamefont {P.~M.}\ \bibnamefont
  {Campbell}}, \bibinfo {author} {\bibfnamefont {J.~C.}\ \bibnamefont
  {Culbertson}}, \bibinfo {author} {\bibfnamefont {E.}~\bibnamefont {Aifer}},
  \bibinfo {author} {\bibfnamefont {F.~K.}\ \bibnamefont {Perkins}}, \bibinfo
  {author} {\bibfnamefont {R.~L.}\ \bibnamefont {Myers-Ward}}, \bibinfo
  {author} {\bibfnamefont {J.~K.}\ \bibnamefont {Hite}}, \bibinfo {author}
  {\bibfnamefont {C.~R.}\ \bibnamefont {Eddy~Jr}}, \bibinfo {author}
  {\bibfnamefont {G.~G.}\ \bibnamefont {Jernigan}},  \emph {et~al.},\
  }\href@noop {} {\bibfield  {journal} {\bibinfo  {journal} {Nano Lett.}\
  }\textbf {\bibinfo {volume} {10}},\ \bibinfo {pages} {3962} (\bibinfo {year}
  {2010})}\BibitemShut {NoStop}%
\bibitem [{\citenamefont {Wang}\ \emph
  {et~al.}(2014{\natexlab{a}})\citenamefont {Wang}, \citenamefont {Gao},
  \citenamefont {Li}, \citenamefont {Lin}, \citenamefont {Li}, \citenamefont
  {Yu},\ and\ \citenamefont {Feng}}]{wang2014classical}%
  \BibitemOpen
  \bibfield  {author} {\bibinfo {author} {\bibfnamefont {W.}~\bibnamefont
  {Wang}}, \bibinfo {author} {\bibfnamefont {K.}~\bibnamefont {Gao}}, \bibinfo
  {author} {\bibfnamefont {Z.}~\bibnamefont {Li}}, \bibinfo {author}
  {\bibfnamefont {T.}~\bibnamefont {Lin}}, \bibinfo {author} {\bibfnamefont
  {J.}~\bibnamefont {Li}}, \bibinfo {author} {\bibfnamefont {C.}~\bibnamefont
  {Yu}}, \ and\ \bibinfo {author} {\bibfnamefont {Z.}~\bibnamefont {Feng}},\
  }\href@noop {} {\bibfield  {journal} {\bibinfo  {journal} {Appl. Phys.
  Lett.}\ }\textbf {\bibinfo {volume} {105}},\ \bibinfo {pages} {182102}
  (\bibinfo {year} {2014}{\natexlab{a}})}\BibitemShut {NoStop}%
\bibitem [{\citenamefont {Kisslinger}\ \emph {et~al.}(2015)\citenamefont
  {Kisslinger}, \citenamefont {Ott}, \citenamefont {Heide}, \citenamefont
  {Kampert}, \citenamefont {Butz}, \citenamefont {Spiecker}, \citenamefont
  {Shallcross},\ and\ \citenamefont {Weber}}]{kisslinger2015linear}%
  \BibitemOpen
  \bibfield  {author} {\bibinfo {author} {\bibfnamefont {F.}~\bibnamefont
  {Kisslinger}}, \bibinfo {author} {\bibfnamefont {C.}~\bibnamefont {Ott}},
  \bibinfo {author} {\bibfnamefont {C.}~\bibnamefont {Heide}}, \bibinfo
  {author} {\bibfnamefont {E.}~\bibnamefont {Kampert}}, \bibinfo {author}
  {\bibfnamefont {B.}~\bibnamefont {Butz}}, \bibinfo {author} {\bibfnamefont
  {E.}~\bibnamefont {Spiecker}}, \bibinfo {author} {\bibfnamefont
  {S.}~\bibnamefont {Shallcross}}, \ and\ \bibinfo {author} {\bibfnamefont
  {H.~B.}\ \bibnamefont {Weber}},\ }\href@noop {} {\bibfield  {journal}
  {\bibinfo  {journal} {Nat. Phys.}\ }\textbf {\bibinfo {volume} {11}},\
  \bibinfo {pages} {650} (\bibinfo {year} {2015})}\BibitemShut {NoStop}%
\bibitem [{\citenamefont {Vasileva}\ \emph {et~al.}(2016)\citenamefont
  {Vasileva}, \citenamefont {Smirnov}, \citenamefont {Ivanov}, \citenamefont
  {Vasilyev}, \citenamefont {Alekseev}, \citenamefont {Dmitriev}, \citenamefont
  {Gornyi}, \citenamefont {Kachorovskii}, \citenamefont {Titov}, \citenamefont
  {Narozhny},\ and\ \citenamefont {Haug}}]{PhysRevB.93.195430}%
  \BibitemOpen
  \bibfield  {author} {\bibinfo {author} {\bibfnamefont {G.~Y.}\ \bibnamefont
  {Vasileva}}, \bibinfo {author} {\bibfnamefont {D.}~\bibnamefont {Smirnov}},
  \bibinfo {author} {\bibfnamefont {Y.~L.}\ \bibnamefont {Ivanov}}, \bibinfo
  {author} {\bibfnamefont {Y.~B.}\ \bibnamefont {Vasilyev}}, \bibinfo {author}
  {\bibfnamefont {P.~S.}\ \bibnamefont {Alekseev}}, \bibinfo {author}
  {\bibfnamefont {A.~P.}\ \bibnamefont {Dmitriev}}, \bibinfo {author}
  {\bibfnamefont {I.~V.}\ \bibnamefont {Gornyi}}, \bibinfo {author}
  {\bibfnamefont {V.~Y.}\ \bibnamefont {Kachorovskii}}, \bibinfo {author}
  {\bibfnamefont {M.}~\bibnamefont {Titov}}, \bibinfo {author} {\bibfnamefont
  {B.~N.}\ \bibnamefont {Narozhny}}, \ and\ \bibinfo {author} {\bibfnamefont
  {R.~J.}\ \bibnamefont {Haug}},\ }\href {\doibase 10.1103/PhysRevB.93.195430}
  {\bibfield  {journal} {\bibinfo  {journal} {Phys. Rev. B}\ }\textbf {\bibinfo
  {volume} {93}},\ \bibinfo {pages} {195430} (\bibinfo {year}
  {2016})}\BibitemShut {NoStop}%
\bibitem [{\citenamefont {Wang}\ \emph
  {et~al.}(2014{\natexlab{b}})\citenamefont {Wang}, \citenamefont {Yang},
  \citenamefont {Li}, \citenamefont {Zhao}, \citenamefont {Wang}, \citenamefont
  {Zhang},\ and\ \citenamefont {Gao}}]{wang2014granularity}%
  \BibitemOpen
  \bibfield  {author} {\bibinfo {author} {\bibfnamefont {Z.}~\bibnamefont
  {Wang}}, \bibinfo {author} {\bibfnamefont {L.}~\bibnamefont {Yang}}, \bibinfo
  {author} {\bibfnamefont {X.}~\bibnamefont {Li}}, \bibinfo {author}
  {\bibfnamefont {X.}~\bibnamefont {Zhao}}, \bibinfo {author} {\bibfnamefont
  {H.}~\bibnamefont {Wang}}, \bibinfo {author} {\bibfnamefont {Z.}~\bibnamefont
  {Zhang}}, \ and\ \bibinfo {author} {\bibfnamefont {X.~P.}\ \bibnamefont
  {Gao}},\ }\href@noop {} {\bibfield  {journal} {\bibinfo  {journal} {Nano
  Lett.}\ }\textbf {\bibinfo {volume} {14}},\ \bibinfo {pages} {6510} (\bibinfo
  {year} {2014}{\natexlab{b}})}\BibitemShut {NoStop}%
\bibitem [{\citenamefont {Yan}\ \emph {et~al.}(2013)\citenamefont {Yan},
  \citenamefont {Wang}, \citenamefont {Yu},\ and\ \citenamefont
  {Liao}}]{yan2013large}%
  \BibitemOpen
  \bibfield  {author} {\bibinfo {author} {\bibfnamefont {Y.}~\bibnamefont
  {Yan}}, \bibinfo {author} {\bibfnamefont {L.-X.}\ \bibnamefont {Wang}},
  \bibinfo {author} {\bibfnamefont {D.-P.}\ \bibnamefont {Yu}}, \ and\ \bibinfo
  {author} {\bibfnamefont {Z.-M.}\ \bibnamefont {Liao}},\ }\href@noop {}
  {\bibfield  {journal} {\bibinfo  {journal} {Appl. Phys. Lett.}\ }\textbf
  {\bibinfo {volume} {103}},\ \bibinfo {pages} {033106} (\bibinfo {year}
  {2013})}\BibitemShut {NoStop}%
\bibitem [{\citenamefont {Wang}\ \emph
  {et~al.}(2012{\natexlab{c}})\citenamefont {Wang}, \citenamefont {Du},
  \citenamefont {Dou},\ and\ \citenamefont {Zhang}}]{PhysRevLett.108.266806}%
  \BibitemOpen
  \bibfield  {author} {\bibinfo {author} {\bibfnamefont {X.}~\bibnamefont
  {Wang}}, \bibinfo {author} {\bibfnamefont {Y.}~\bibnamefont {Du}}, \bibinfo
  {author} {\bibfnamefont {S.}~\bibnamefont {Dou}}, \ and\ \bibinfo {author}
  {\bibfnamefont {C.}~\bibnamefont {Zhang}},\ }\href {\doibase
  10.1103/PhysRevLett.108.266806} {\bibfield  {journal} {\bibinfo  {journal}
  {Phys. Rev. Lett.}\ }\textbf {\bibinfo {volume} {108}},\ \bibinfo {pages}
  {266806} (\bibinfo {year} {2012}{\natexlab{c}})}\BibitemShut {NoStop}%
\bibitem [{\citenamefont {Tang}\ \emph {et~al.}(2011)\citenamefont {Tang},
  \citenamefont {Liang}, \citenamefont {Qiu},\ and\ \citenamefont
  {Gao}}]{tang2011two}%
  \BibitemOpen
  \bibfield  {author} {\bibinfo {author} {\bibfnamefont {H.}~\bibnamefont
  {Tang}}, \bibinfo {author} {\bibfnamefont {D.}~\bibnamefont {Liang}},
  \bibinfo {author} {\bibfnamefont {R.~L.}\ \bibnamefont {Qiu}}, \ and\
  \bibinfo {author} {\bibfnamefont {X.~P.}\ \bibnamefont {Gao}},\ }\href@noop
  {} {\bibfield  {journal} {\bibinfo  {journal} {ACS Nano}\ }\textbf {\bibinfo
  {volume} {5}},\ \bibinfo {pages} {7510} (\bibinfo {year} {2011})}\BibitemShut
  {NoStop}%
\bibitem [{\citenamefont {He}\ \emph {et~al.}(2013)\citenamefont {He},
  \citenamefont {Liu}, \citenamefont {Li}, \citenamefont {Guo}, \citenamefont
  {Xu}, \citenamefont {Xie},\ and\ \citenamefont {Wang}}]{he2013disorder}%
  \BibitemOpen
  \bibfield  {author} {\bibinfo {author} {\bibfnamefont {H.}~\bibnamefont
  {He}}, \bibinfo {author} {\bibfnamefont {H.}~\bibnamefont {Liu}}, \bibinfo
  {author} {\bibfnamefont {B.}~\bibnamefont {Li}}, \bibinfo {author}
  {\bibfnamefont {X.}~\bibnamefont {Guo}}, \bibinfo {author} {\bibfnamefont
  {Z.}~\bibnamefont {Xu}}, \bibinfo {author} {\bibfnamefont {M.}~\bibnamefont
  {Xie}}, \ and\ \bibinfo {author} {\bibfnamefont {J.}~\bibnamefont {Wang}},\
  }\href@noop {} {\bibfield  {journal} {\bibinfo  {journal} {Appl. Phys.
  Lett.}\ }\textbf {\bibinfo {volume} {103}},\ \bibinfo {pages} {031606}
  (\bibinfo {year} {2013})}\BibitemShut {NoStop}%
\bibitem [{\citenamefont {Abrikosov}(1998)}]{PhysRevB.58.2788}%
  \BibitemOpen
  \bibfield  {author} {\bibinfo {author} {\bibfnamefont {A.~A.}\ \bibnamefont
  {Abrikosov}},\ }\href {\doibase 10.1103/PhysRevB.58.2788} {\bibfield
  {journal} {\bibinfo  {journal} {Phys. Rev. B}\ }\textbf {\bibinfo {volume}
  {58}},\ \bibinfo {pages} {2788} (\bibinfo {year} {1998})}\BibitemShut
  {NoStop}%
\bibitem [{\citenamefont {Parish}\ and\ \citenamefont
  {Littlewood}(2003)}]{parish2003non}%
  \BibitemOpen
  \bibfield  {author} {\bibinfo {author} {\bibfnamefont {M.}~\bibnamefont
  {Parish}}\ and\ \bibinfo {author} {\bibfnamefont {P.}~\bibnamefont
  {Littlewood}},\ }\href@noop {} {\bibfield  {journal} {\bibinfo  {journal}
  {Nature}\ }\textbf {\bibinfo {volume} {426}},\ \bibinfo {pages} {162}
  (\bibinfo {year} {2003})}\BibitemShut {NoStop}%
\bibitem [{\citenamefont {Hu}\ \emph {et~al.}(2007)\citenamefont {Hu},
  \citenamefont {Parish},\ and\ \citenamefont
  {Rosenbaum}}]{PhysRevB.75.214203}%
  \BibitemOpen
  \bibfield  {author} {\bibinfo {author} {\bibfnamefont {J.}~\bibnamefont
  {Hu}}, \bibinfo {author} {\bibfnamefont {M.~M.}\ \bibnamefont {Parish}}, \
  and\ \bibinfo {author} {\bibfnamefont {T.~F.}\ \bibnamefont {Rosenbaum}},\
  }\href {\doibase 10.1103/PhysRevB.75.214203} {\bibfield  {journal} {\bibinfo
  {journal} {Phys. Rev. B}\ }\textbf {\bibinfo {volume} {75}},\ \bibinfo
  {pages} {214203} (\bibinfo {year} {2007})}\BibitemShut {NoStop}%
\bibitem [{\citenamefont {Kisslinger}\ \emph {et~al.}(2017)\citenamefont
  {Kisslinger}, \citenamefont {Ott},\ and\ \citenamefont
  {Weber}}]{PhysRevB.95.024204}%
  \BibitemOpen
  \bibfield  {author} {\bibinfo {author} {\bibfnamefont {F.}~\bibnamefont
  {Kisslinger}}, \bibinfo {author} {\bibfnamefont {C.}~\bibnamefont {Ott}}, \
  and\ \bibinfo {author} {\bibfnamefont {H.~B.}\ \bibnamefont {Weber}},\ }\href
  {\doibase 10.1103/PhysRevB.95.024204} {\bibfield  {journal} {\bibinfo
  {journal} {Phys. Rev. B}\ }\textbf {\bibinfo {volume} {95}},\ \bibinfo
  {pages} {024204} (\bibinfo {year} {2017})}\BibitemShut {NoStop}%
\bibitem [{\citenamefont {Alekseev}\ \emph {et~al.}(2015)\citenamefont
  {Alekseev}, \citenamefont {Dmitriev}, \citenamefont {Gornyi}, \citenamefont
  {Kachorovskii}, \citenamefont {Narozhny}, \citenamefont {Sch\"utt},\ and\
  \citenamefont {Titov}}]{PhysRevLett.114.156601}%
  \BibitemOpen
  \bibfield  {author} {\bibinfo {author} {\bibfnamefont {P.~S.}\ \bibnamefont
  {Alekseev}}, \bibinfo {author} {\bibfnamefont {A.~P.}\ \bibnamefont
  {Dmitriev}}, \bibinfo {author} {\bibfnamefont {I.~V.}\ \bibnamefont
  {Gornyi}}, \bibinfo {author} {\bibfnamefont {V.~Y.}\ \bibnamefont
  {Kachorovskii}}, \bibinfo {author} {\bibfnamefont {B.~N.}\ \bibnamefont
  {Narozhny}}, \bibinfo {author} {\bibfnamefont {M.}~\bibnamefont {Sch\"utt}},
  \ and\ \bibinfo {author} {\bibfnamefont {M.}~\bibnamefont {Titov}},\ }\href
  {\doibase 10.1103/PhysRevLett.114.156601} {\bibfield  {journal} {\bibinfo
  {journal} {Phys. Rev. Lett.}\ }\textbf {\bibinfo {volume} {114}},\ \bibinfo
  {pages} {156601} (\bibinfo {year} {2015})}\BibitemShut {NoStop}%
\bibitem [{\citenamefont {Alekseev}\ \emph {et~al.}(2017)\citenamefont
  {Alekseev}, \citenamefont {Dmitriev}, \citenamefont {Gornyi}, \citenamefont
  {Kachorovskii}, \citenamefont {Narozhny}, \citenamefont {Sch\"utt},\ and\
  \citenamefont {Titov}}]{alekseev2016magnetoresistance}%
  \BibitemOpen
  \bibfield  {author} {\bibinfo {author} {\bibfnamefont {P.~S.}\ \bibnamefont
  {Alekseev}}, \bibinfo {author} {\bibfnamefont {A.~P.}\ \bibnamefont
  {Dmitriev}}, \bibinfo {author} {\bibfnamefont {I.~V.}\ \bibnamefont
  {Gornyi}}, \bibinfo {author} {\bibfnamefont {V.~Y.}\ \bibnamefont
  {Kachorovskii}}, \bibinfo {author} {\bibfnamefont {B.~N.}\ \bibnamefont
  {Narozhny}}, \bibinfo {author} {\bibfnamefont {M.}~\bibnamefont {Sch\"utt}},
  \ and\ \bibinfo {author} {\bibfnamefont {M.}~\bibnamefont {Titov}},\ }\href
  {\doibase 10.1103/PhysRevB.95.165410} {\bibfield  {journal} {\bibinfo
  {journal} {Phys. Rev. B}\ }\textbf {\bibinfo {volume} {95}},\ \bibinfo
  {pages} {165410} (\bibinfo {year} {2017})}\BibitemShut {NoStop}%
\bibitem [{\citenamefont {Bl\"ochl}(1994)}]{PhysRevB.50.17953}%
  \BibitemOpen
  \bibfield  {author} {\bibinfo {author} {\bibfnamefont {P.~E.}\ \bibnamefont
  {Bl\"ochl}},\ }\href@noop {} {\bibfield  {journal} {\bibinfo  {journal}
  {Phys. Rev. B}\ }\textbf {\bibinfo {volume} {50}},\ \bibinfo {pages} {17953}
  (\bibinfo {year} {1994})}\BibitemShut {NoStop}%
\bibitem [{\citenamefont {Kresse}\ and\ \citenamefont
  {Joubert}(1999)}]{PhysRevB.59.1758}%
  \BibitemOpen
  \bibfield  {author} {\bibinfo {author} {\bibfnamefont {G.}~\bibnamefont
  {Kresse}}\ and\ \bibinfo {author} {\bibfnamefont {D.}~\bibnamefont
  {Joubert}},\ }\href@noop {} {\bibfield  {journal} {\bibinfo  {journal} {Phys.
  Rev. B}\ }\textbf {\bibinfo {volume} {59}},\ \bibinfo {pages} {1758}
  (\bibinfo {year} {1999})}\BibitemShut {NoStop}%
\bibitem [{\citenamefont {Kresse}\ and\ \citenamefont
  {Hafner}(1993)}]{PhysRevB.47.558}%
  \BibitemOpen
  \bibfield  {author} {\bibinfo {author} {\bibfnamefont {G.}~\bibnamefont
  {Kresse}}\ and\ \bibinfo {author} {\bibfnamefont {J.}~\bibnamefont
  {Hafner}},\ }\href@noop {} {\bibfield  {journal} {\bibinfo  {journal} {Phys.
  Rev. B}\ }\textbf {\bibinfo {volume} {47}},\ \bibinfo {pages} {558} (\bibinfo
  {year} {1993})}\BibitemShut {NoStop}%
\bibitem [{\citenamefont {Kresse}\ and\ \citenamefont
  {Furthm{\"u}ller}(1996)}]{kresse1996efficiency}%
  \BibitemOpen
  \bibfield  {author} {\bibinfo {author} {\bibfnamefont {G.}~\bibnamefont
  {Kresse}}\ and\ \bibinfo {author} {\bibfnamefont {J.}~\bibnamefont
  {Furthm{\"u}ller}},\ }\href@noop {} {\bibfield  {journal} {\bibinfo
  {journal} {Comp. Mater. Sci.}\ }\textbf {\bibinfo {volume} {6}},\ \bibinfo
  {pages} {15} (\bibinfo {year} {1996})}\BibitemShut {NoStop}%
\bibitem [{\citenamefont {Kresse}\ and\ \citenamefont
  {Furthm\"uller}(1996)}]{PhysRevB.54.11169}%
  \BibitemOpen
  \bibfield  {author} {\bibinfo {author} {\bibfnamefont {G.}~\bibnamefont
  {Kresse}}\ and\ \bibinfo {author} {\bibfnamefont {J.}~\bibnamefont
  {Furthm\"uller}},\ }\href@noop {} {\bibfield  {journal} {\bibinfo  {journal}
  {Phys. Rev. B}\ }\textbf {\bibinfo {volume} {54}},\ \bibinfo {pages} {11169}
  (\bibinfo {year} {1996})}\BibitemShut {NoStop}%
\bibitem [{\citenamefont {Perdew}\ \emph {et~al.}(1996)\citenamefont {Perdew},
  \citenamefont {Burke},\ and\ \citenamefont
  {Ernzerhof}}]{PhysRevLett.77.3865}%
  \BibitemOpen
  \bibfield  {author} {\bibinfo {author} {\bibfnamefont {J.~P.}\ \bibnamefont
  {Perdew}}, \bibinfo {author} {\bibfnamefont {K.}~\bibnamefont {Burke}}, \
  and\ \bibinfo {author} {\bibfnamefont {M.}~\bibnamefont {Ernzerhof}},\
  }\href@noop {} {\bibfield  {journal} {\bibinfo  {journal} {Phys. Rev. Lett.}\
  }\textbf {\bibinfo {volume} {77}},\ \bibinfo {pages} {3865} (\bibinfo {year}
  {1996})}\BibitemShut {NoStop}%
\bibitem [{\citenamefont {Marzari}\ and\ \citenamefont
  {Vanderbilt}(1997)}]{PhysRevB.56.12847}%
  \BibitemOpen
  \bibfield  {author} {\bibinfo {author} {\bibfnamefont {N.}~\bibnamefont
  {Marzari}}\ and\ \bibinfo {author} {\bibfnamefont {D.}~\bibnamefont
  {Vanderbilt}},\ }\href@noop {} {\bibfield  {journal} {\bibinfo  {journal}
  {Phys. Rev. B}\ }\textbf {\bibinfo {volume} {56}},\ \bibinfo {pages} {12847}
  (\bibinfo {year} {1997})}\BibitemShut {NoStop}%
\bibitem [{\citenamefont {Souza}\ \emph {et~al.}(2001)\citenamefont {Souza},
  \citenamefont {Marzari},\ and\ \citenamefont
  {Vanderbilt}}]{PhysRevB.65.035109}%
  \BibitemOpen
  \bibfield  {author} {\bibinfo {author} {\bibfnamefont {I.}~\bibnamefont
  {Souza}}, \bibinfo {author} {\bibfnamefont {N.}~\bibnamefont {Marzari}}, \
  and\ \bibinfo {author} {\bibfnamefont {D.}~\bibnamefont {Vanderbilt}},\
  }\href@noop {} {\bibfield  {journal} {\bibinfo  {journal} {Phys. Rev. B}\
  }\textbf {\bibinfo {volume} {65}},\ \bibinfo {pages} {035109} (\bibinfo
  {year} {2001})}\BibitemShut {NoStop}%
\bibitem [{\citenamefont {Brechtel}\ \emph {et~al.}(1981)\citenamefont
  {Brechtel}, \citenamefont {Cordier},\ and\ \citenamefont
  {Sch{\"a}fer}}]{brechtel1981neue}%
  \BibitemOpen
  \bibfield  {author} {\bibinfo {author} {\bibfnamefont {E.}~\bibnamefont
  {Brechtel}}, \bibinfo {author} {\bibfnamefont {G.}~\bibnamefont {Cordier}}, \
  and\ \bibinfo {author} {\bibfnamefont {H.}~\bibnamefont {Sch{\"a}fer}},\
  }\href@noop {} {\bibfield  {journal} {\bibinfo  {journal} {J. Less-Common
  Met.}\ }\textbf {\bibinfo {volume} {79}},\ \bibinfo {pages} {131} (\bibinfo
  {year} {1981})}\BibitemShut {NoStop}%
\bibitem [{\citenamefont {McKenzie}\ \emph {et~al.}(1998)\citenamefont
  {McKenzie}, \citenamefont {Qualls}, \citenamefont {Han},\ and\ \citenamefont
  {Brooks}}]{PhysRevB.57.11854}%
  \BibitemOpen
  \bibfield  {author} {\bibinfo {author} {\bibfnamefont {R.~H.}\ \bibnamefont
  {McKenzie}}, \bibinfo {author} {\bibfnamefont {J.~S.}\ \bibnamefont
  {Qualls}}, \bibinfo {author} {\bibfnamefont {S.~Y.}\ \bibnamefont {Han}}, \
  and\ \bibinfo {author} {\bibfnamefont {J.~S.}\ \bibnamefont {Brooks}},\
  }\href {\doibase 10.1103/PhysRevB.57.11854} {\bibfield  {journal} {\bibinfo
  {journal} {Phys. Rev. B}\ }\textbf {\bibinfo {volume} {57}},\ \bibinfo
  {pages} {11854} (\bibinfo {year} {1998})}\BibitemShut {NoStop}%
\bibitem [{\citenamefont {Guo}\ \emph {et~al.}(2016)\citenamefont {Guo},
  \citenamefont {Yang}, \citenamefont {Zhang}, \citenamefont {Liu},\ and\
  \citenamefont {Lu}}]{PhysRevB.93.235142}%
  \BibitemOpen
  \bibfield  {author} {\bibinfo {author} {\bibfnamefont {P.-J.}\ \bibnamefont
  {Guo}}, \bibinfo {author} {\bibfnamefont {H.-C.}\ \bibnamefont {Yang}},
  \bibinfo {author} {\bibfnamefont {B.-J.}\ \bibnamefont {Zhang}}, \bibinfo
  {author} {\bibfnamefont {K.}~\bibnamefont {Liu}}, \ and\ \bibinfo {author}
  {\bibfnamefont {Z.-Y.}\ \bibnamefont {Lu}},\ }\href {\doibase
  10.1103/PhysRevB.93.235142} {\bibfield  {journal} {\bibinfo  {journal} {Phys.
  Rev. B}\ }\textbf {\bibinfo {volume} {93}},\ \bibinfo {pages} {235142}
  (\bibinfo {year} {2016})}\BibitemShut {NoStop}%
\bibitem [{\citenamefont {Kozlova}\ \emph {et~al.}(2012)\citenamefont
  {Kozlova}, \citenamefont {Mori}, \citenamefont {Makarovsky}, \citenamefont
  {Eaves}, \citenamefont {Zhuang}, \citenamefont {Krier},\ and\ \citenamefont
  {Patan{\`e}}}]{kozlova2012linear}%
  \BibitemOpen
  \bibfield  {author} {\bibinfo {author} {\bibfnamefont {N.}~\bibnamefont
  {Kozlova}}, \bibinfo {author} {\bibfnamefont {N.}~\bibnamefont {Mori}},
  \bibinfo {author} {\bibfnamefont {O.}~\bibnamefont {Makarovsky}}, \bibinfo
  {author} {\bibfnamefont {L.}~\bibnamefont {Eaves}}, \bibinfo {author}
  {\bibfnamefont {Q.}~\bibnamefont {Zhuang}}, \bibinfo {author} {\bibfnamefont
  {A.}~\bibnamefont {Krier}}, \ and\ \bibinfo {author} {\bibfnamefont
  {A.}~\bibnamefont {Patan{\`e}}},\ }\href@noop {} {\bibfield  {journal}
  {\bibinfo  {journal} {Nat. Commun.}\ }\textbf {\bibinfo {volume} {3}},\
  \bibinfo {pages} {1097} (\bibinfo {year} {2012})}\BibitemShut {NoStop}%
\bibitem [{\citenamefont {Kokalj}(2003)}]{kokalj2003computer}%
  \BibitemOpen
  \bibfield  {author} {\bibinfo {author} {\bibfnamefont {A.}~\bibnamefont
  {Kokalj}},\ }\href@noop {} {\bibfield  {journal} {\bibinfo  {journal} {Comp.
  Mater. Sci.}\ }\textbf {\bibinfo {volume} {28}},\ \bibinfo {pages} {155}
  (\bibinfo {year} {2003})}\BibitemShut {NoStop}%
\end{thebibliography}%
\end{document}